\documentclass[11pt]{article}

\usepackage{amssymb,amsthm}
\usepackage[totalwidth=17cm,totalheight=24cm]{geometry}
\newcommand{\C}{{\mathbb C}}
\newcommand{\R}{{\mathbb R}}
\newcommand{\im}{{\mathrm i}}
\newcommand{\A}{{\mathcal A}}
\newcommand{\cH}{{\mathcal H}}

\newcommand\be{\begin{eqnarray}}
\newcommand\ee{\end{eqnarray}}

\begin{document}

\title{Gravity as a diffeomorphism invariant gauge theory}
\author{Kirill Krasnov \\
{\it School of Mathematical Sciences} \\ {\it University of Nottingham}}
  
\date{v3: May 2011}
\maketitle

\begin{abstract} A general diffeomorphism invariant ${\rm SU}(2)$ gauge theory is a gravity theory with two propagating polarizations of the graviton. We develop this description of gravity, in particular for future applications to the perturbative quantization. Thus, the linearized theory, gauge symmetries, gauge fixing are discussed in detail, and the propagator is obtained. The propagator takes a simple form of that of Yang-Mills theory with an additional projector on diffeomorphism equivalence classes of connections inserted. In our approach the gravitational perturbation theory takes a rather unusual form in that the Planck length is no longer fundamental.
\end{abstract}

\section{Introduction}

The field theory approach to gravity, see e.g.  \cite{Feynman:1996kb}, tells us that gravity is not a gauge theory. Indeed, the carriers of force in a gauge theory (such as e.g. Maxwell electrodynamics) are spin one particles. For this reason there are two types of charged objects interacting by exchange of carriers of force - those negatively and positively charged. Like particles repel and unlikes attract. In contrast, there is only one type of charge in gravity and everything attracts everything. Thus, gravity is not a gauge theory, see \cite{Feynman:1996kb} for a more detailed discussion.

This simple argument forbids a direct gauge theory description of gravity. It says nothing, however, about less direct possible relationships. And indeed, a relation of a completely different type is now being very popular. This has its origin in the open-closed string duality, which implies that amplitudes for closed strings are squares of those for open strings. Since the low energy limit of the closed string theory is gravity, and that for open strings is gauge theory, this implies that scattering amplitudes for gravitons must be expressible as squares of amplitudes for gluons, see e.g. a review \cite{Bern:2002kj} and/or a more recent paper \cite{Bern:2010yg} and references therein. The relationship is not direct, and it is in particular not easy to find a Lagrangian version of the correspondence. However, it has recently led to some very interesting developments on loop divergences in $N=8$ supergravity. Another example of a gauge theory/gravity relation is the AdS/CFT correspondence \cite{Witten:1998qj} of string theory.

The aim of the present paper is to develop (further, see historical remarks below) yet another gravity/gauge theory correspondence. Currently there appears to be no relation between the present story and that of \cite{Bern:2002kj}. The relationship of interest for us here has its origins in the discovery of Plebanski \cite{Plebanski:1977zz} that certain triple of self-dual two-forms can be used as the basic variables for gravity\footnote{Similar ideas has appeared in the literature much earlier, see e.g. \cite{Krasnov:2009pu} for historical remarks, but it was Plebanski who proposed to reformulate general relativity without the metric, with only two-forms as dynamical variables.}. The same "self-dual" formulation of general relativity (GR) has been rediscovered a decade later by Ashtekar \cite{Ashtekar:1987gu} via a completely different path of a canonical transformation on the phase space of GR. The two discoveries were later linked in \cite{Jacobson:1988yy}, and the outcome was a realisation that gravity can be reformulated as a theory whose phase space coincides with that of an ${\rm SU}(2)$ gauge theory. This gravity/gauge theory relationship was taken one step further in \cite{Capovilla:1989ac}. Thus, it was realized that the two-form fields of Plebanski formulation of GR \cite{Plebanski:1977zz} can be integrated out to obtain a "pure connection" formulation of general relativity, where the only dynamical field is an ${\rm SU}(2)$ connection. The result was a completely new perspective on general relativity, in which GR becomes reformulated as a novel type of a theory of the gauge field --- a {\it diffeomorphism invariant gauge theory}. 

The work on "pure connection" formulation of GR \cite{Capovilla:1989ac} has led to some further advances in that it was realized in \cite{Bengtsson:1990qg} that there is not a single diffeomorphism invariant gauge theory, but an infinite parameter class of them. All these theories share the same key properties with GR, as they have the same number of propagating degrees of freedom (DOF). Thus, for any theory in the class the phase space is that of an ${\rm SU}(2)$ gauge theory. However, in addition to the usual ${\rm SU}(2)$ gauge rotations, there are also diffeomorphisms acting on the phase space variables, which reduce the number of propagating DOF from 6 of ${\rm SU}(2)$ gauge theory to 2 of GR. 

Unfortunately, the new "pure connection" viewpoint on GR originating in \cite{Capovilla:1989ac} (and having its roots in Plebanski's key insight \cite{Plebanski:1977zz}) has not been significantly developed. The phase space version \cite{Ashtekar:1987gu} of this story has formed the foundation of the approach of loop quantum gravity \cite{Rovelli:2008zza}, but the pure connection formulation of GR \cite{Capovilla:1989ac} and of the infinite-parameter family \cite{Bengtsson:1990qg} of "neighbours of GR" has not had any significant applications, as far as the author is aware. The main aim of this paper is to revisit this "pure connection" formalism for gravity and develop it further. Our main motivation is a (future) application of this formalism to the perturbative quantization of gravity. However, as it will become clear below, the pure connection perspective on gravity developed here may have other uses. 

We motivate our interest in this formalism for (quantum) gravity with some historical remarks. Thus, the author's interest in the subject started in \cite{Krasnov:2006du} from a simple power counting argument describing how the non-renormalizability of GR manifests itself in the Plebanski formulation \cite{Plebanski:1977zz}. The outcome was an infinite-parameter family of Plebanski-like theories, where the constraint term of the Plebanski action was replaced by a "potential" term for the would-be Lagrange multipliers. Each of the new theories is just the familiar from discussions of non-renormalizability counterterm corrected GR (in disguise), and so the interpretation of the infinite number of new parameters is that they are related to coefficients in front of counterterms constructed from the curvature and its derivatives in the usual metric description of gravity.  It was very quickly realized \cite{Bengtsson:2007zzd} that the new infinite-parameter family of theories \cite{Krasnov:2006du} is essentially the same as the one introduced and studied by Begtsson and collaborators a decade earlier \cite{Bengtsson:1990qg}, with the difference being that \cite{Bengtsson:1990qg} worked at the level of a "pure connection" formulation, while the theories \cite{Krasnov:2006du} are formulated as Plebanski-like theories with two-form fields as the basic variables. 

The class of gravity theories \cite{Bengtsson:1990qg}, \cite{Krasnov:2006du} can be thought of as summing at least some of the quantum corrections that arise in the process of renormalization of GR, and in \cite{Krasnov:2009ik} this was confirmed by directly exhibiting the familiar GR counterterms as appearing from \cite{Krasnov:2006du}. The work \cite{Krasnov:2006du} also conjectured that this class of gravity theories sums up {\it all} the arising quantum corrections; in other words, it was conjectured that the class \cite{Krasnov:2006du} is closed under the renormalization, and that the arising renormalization group flow is that in the space of "potential" functions defining the theory. 

At the time of writing \cite{Krasnov:2006du} the only motivation for this conjecture was the author's optimism --- the conjecture did not contradict anything one knew about the non-renormalizability of GR, and was the most optimistic scenario for how the divergences of GR might organise themselves. The remark \cite{Bengtsson:2007zzd} relating the Plebanski-like theories \cite{Krasnov:2006du} to the pure connection theories \cite{Bengtsson:1990qg} brought with it an additional justification. Thus, a closer look at these theories made it clear that they are just the most general diffeomorphism invariant gauge theories. The class of such theories should therefore be closed under the renormalization, because any counterterm that can be needed for cancelling the arising quantum divergences is already included into the action, see \cite{Krasnov:2009ip} for the first spell-out of this argument.

One of the aims of the present paper is to set the stage for a systematic study of the quantum perturbation theory for the gravitational theories introduced in \cite{Krasnov:2006du} (and previously in \cite{Bengtsson:1990qg}). Our final goal is to settle the status of the conjecture of \cite{Krasnov:2006du} that these class of theories is closed under the renormalization, and then to compute the resulting renormalization group flow. However, it would be impractical to try to write up all the necessary calculations in a single paper. For this reason, in the present paper we develop the classical theory to the extent that the propagating degrees of freedom (gravitons) are manifest. We also do some preliminary steps necessary for the perturbative loop computations in that the gauge fixing is discussed in details and the propagator is obtained. It is then straightforward to start to compute loop diagrams. This is however not attempted in the present paper, and the task to develop a sufficiently economical way to study the renormalization is left to future work.

Apart from just setting up the stage for future quantum calculations, a somewhat unexpected outcome of this work is a completely new viewpoint on the gravitational perturbation theory. As we shall see, in the present diffeomorphism invariant gauge theoretic approach to gravity, the fundamental scale is set not by the Newton's constant, which does not appear in the original formulation of the theory at all, but rather by the radius of curvature of the background that is used to expand the theory around. Thus, the natural fundamental length scale is set by the cosmological constant. This has the effect that in our theory the Newton's constant becomes a derived quantity. This leads to some puzzles about the cutoff scale of our perturbation theory, to be discussed towards the end of the paper. Another point that is worth emphasizing from the outset is that our gauge-theoretic approach to gravity only works for a non-zero value of the cosmological constant $\Lambda$. As we shall see more explicitly below, the actions we work with blows up in the limit $\Lambda\to 0$. The puzzles about the behaviour of the perturbation theory that we discuss in the main text are directly related to this feature. 

To summarize, the main aim of this work is to develop a new approach to the gravitational perturbation theory, for future use in particular in the quantum loop calculations. What makes this paper distinct from previous works (in particular of this author) is that here for the first time the "pure connection" formalism close in spirit to the formulation in \cite{Bengtsson:1990qg} is used as a starting point for the gravitational perturbation theory. Thus, all previous works on theories \cite{Krasnov:2006du} used the two-form formulation. The gravitational perturbation theory in the two-form formulation is similar to that in the usual metric approach, see \cite{Krasnov:2009ik}. In particular, the fundamental scale that determines the self-coupling of the gravitons and sets the scale of the strong coupling regime is, as in the usual metric case, the Planck scale. However, the number of field components one has to works with in the two-form formulation is quite large --- it is that of an ${\rm SU}(2)$ Lie algebra-valued two-form field. Moreover, there are second class constraints that require the path integral measure to be somewhat non-trivial. For all these reasons it proved to be rather difficult to set up an economical perturbation theory in the two-form formalism. At the same time, for a long time it seemed that the "pure connection" formulation is ill suited for being a starting point of a perturbative description, as it was not at all clear how one can expand the theory around the Minkowski spacetime background which corresponds to a zero connection, see e.g. remarks in \cite{TorresGomez:2009gs}. 

In this paper the prejudices about the "pure connection" formulation of gravity are put aside and this formulation is used as a starting point for the gravitational perturbation theory. And, as we hope to convince the reader, this formulation can be used rather effectively, in that the arising perturbation theory is reasonably economical. In particular, the linearized theory is very simple (arguably simpler than in the metric description), and the propagator can be obtained without too much difficulty. As we shall see, in the "pure connection" formalism developed here gravity becomes not too dissimilar to ${\rm SU}(2)$ gauge theory, the main difference being that a certain additional projector on diffeomorphism equivalence classes is inserted into the standard $1/k^2$ propagator of the gauge theory. This gives hope that the renormalization in this class of gravity theories will eventually become manageable. As we have already mentioned, this is left to the future work.

What is new in this work as compared to previous works on the "pure connection" formulation, in particular the work  \cite{Bengtsson:1990qg} and works by Bengtsson and collaborators that followed, is that our treatment uses in an essential way the formulation in terms of a homogeneous potential function applied to a matrix-valued 4-form. This was developed in earlier works of the author, and first spelled out in \cite{Krasnov:2009iy} for the version of the theory that uses a two-form field, and in \cite{Krasnov:2009ip} for the pure connection formulation. This formulation renders the action principle of the theory very compact, and makes it possible to set up the perturbation theory without too much difficulty.

Before we proceed with a description of the theory, there are a few things that ought to be emphasised to avoid misunderstanding. In our gauge-theoretic approach to gravity the theory (or any of the class of theories that we study) remains as non-renormalizable as GR in the usual metric-based treatment. Thus, as we shall explicitly see below, the coupling constant of our theory has a negative mass dimension, which signals non-renormalizability by power counting. Thus, the final goal of our enterprise is not to show that the theory is renormalizable --- it is not --- but rather to show that the infinite-parameter class of theories that we study is closed under the renormalization, and then to compute the arising renormalization group flow. In other words, we are not after the renormalizability in the usual sense of quantum field theory, which is that a Lagrangian with a finite number of couplings is closed under the renormalization. Rather, we are after the renormalizability in the effective field theory sense of Weinberg, see e.g. \cite{Weinberg:2009bg} for a recent discussion, where any theory is renormalizable once all possible counterterms are added to the action. Our aim is then to show that in the case of gravity in four spacetime dimensions it is sufficient to consider only those counterterms (infinite in number) that can be compactly summed up into our diffeomorphism invariant gauge theory Lagrangian. Should this indeed be the case, the renormalization group flow in the infinite dimensional space of gravity theories will be just a flow in the space of defining functions, and will become manageable. Note once again, however, that the quantum theory, while being our main motivation, is not the subject of the present work.

We would also like to explain at the outset how a gauge theory (with spin one excitations) can describe gravity with its spin two excitations. This is a version of the story "spin one plus spin one is spin two", of relevance for the gauge theory gravity relationship \cite{Bern:2002kj}. There are, however, also significant differences. Thus, the main dynamical field of our theories is an ${\rm SU}(2)$ connection $A_\mu^i$, where $\mu$ is a spacetime index, and $i=1,2,3$ is a Lie algebra one. Let us recall that in the usual gauge theories in Minkowski spacetime the temporal component $A_0^i$ of the connection field becomes a Lagrange multiplier --- the generator of the gauge rotations. Then of the spatial components $A_a^i$, where $a=1,2,3$ is a spatial index, some components are pure gauge in that they can be set to any desired value by a gauge transformation. The physical propagating degrees of freedom of the theory can be described as the gauge equivalence classes of the spatial projection of the connection. In the case of gauge group ${\rm SU}(2)$, the gauge invariance removes 3 of the 9 components of the spatial connection $A_a^i$, leaving two propagating polarizations per each Lie algebra generator. 

As we shall see below, in the case of our gravitational theories the situation is very similar, with the exception that the Lagrangian is in addition invariant under diffeomorphisms. The way this is realized in our theories is that the Lagrangian is simply independent of certain 4 combinations of the connection field $A_\mu^i$. This is where the spin two comes from. Thus, consider once again the spatial projection of the connection $A_a^i$. We shall see that (using the background) it will be possible to identify two types of indices --- the spatial and the internal Lie algebra ones. Once this is done, the spatial connection can be thought of as a $3\times 3$ matrix, or, in representation theoretic terms, it constitutes the spin one tensor spin one representation. This decomposes as spin two plus spin one plus spin zero. On the other hand, the temporal component of the connection $A_0^i$ forms the spin one (adjoint) representation of ${\rm SU}(2)$. The diffeomorphism invariance projects out the spin zero components of the spatial connection $A_a^i$, as well as a certain combination of the spin one component of $A_a^i$ and $A_0^i$, leaving only one of these spin one components in the game. Thus, after the projection induced by the diffeomorphisms, the Lagrangian depends only on the spin two component of $A_a^i$, as well as on the spin one set of Lagrange multipliers --- generators of ${\rm SU}(2)$ rotations. These make the 3 longitudinal components of the 5 component spin two field unphysical, leaving only 2 propagating physical modes. To summarize, in our version of gauge theory/gravity correspondence the spin two also comes from the tensor product of two spin one representations. As in any gauge theory in Minkowski space, one of these spin one representations is supplied by the spatial projection of the connection field. The other spin one is provided by the adjoint representation of the ${\rm SU}(2)$ Lie algebra in which the connection field takes values. 

Our final remark here is about the issues of the reality conditions. As we shall see below, in the physically realistic case of Lorentzian signature gravity, the main dynamical field of our theories is a complex-valued ${\rm SO}(3,\C)$ connection. Thus, appropriate reality conditions need to be imposed to select the field configurations corresponding to real Lorentzian metrics. Our strategy for dealing with these in the present paper is as follows. We shall see that at the level of the linearized theory the reality conditions are straightforward (one can easily determine them from the requirement that the linearized Hamiltonian is positive-definite). In the full theory, however, the reality conditions need to be imposed non-perturbatively. We do not yet know how to do this. However, for many applications, in particular the ones we are most interested in, this is not needed. Thus, for calculations studying the renormalization of gravity (e.g. ones done using the background field method), one performs the Wick rotation to the Riemannian signature metrics. For the later case our gauge field is a real-valued ${\rm SO}(3)$ connection, and the reality conditions are straightforward. Similarly, for the perturbative loop computations one uses the knowledge of the linearized reality conditions to specify the physical external states. It is then only necessary to specify the contour in the complex connections space that is used in the loop integrations. There is typically very little freedom in the choice of this contour provided one wants the integrals to converge. So, again it is possible to perform computations without specifying explicitly the full non-linear theory reality conditions. This state of affairs, is, of course, not completely satisfactory, for one would like to have a complete control over the full theory reality conditions as well. This is, however, beyond the scope of this paper (and is never needed here).

With these preparatory remarks having been made, we can proceed to describe how gravity can be reformulated as a diffeomorphism invariant gauge theory. The organization of the paper is as follows. In Section \ref{sec:formulation} we define an action principle for our theories, explain how their parameterization by a homogenous function works, derive the field equations and verify gauge invariances of the action. Section \ref{sec:backgr} studies the theory linearized around a constant curvature background. In particular, a simple quadratic in the gauge field fluctuations action is obtained, and its Hamiltonian analysis is performed. This confirms the outlined above picture of how the spin two nature of excitations comes about. Section \ref{sec:prop} is central to our analysis. It discusses the gauge-fixing appropriate to the situation at hand, and inverts the gauge-fixed quadratic form to obtain the propagator. In Section \ref{sec:inter} we derive the (cubic and quartic) interaction terms of our theory. We conclude with a brief discussion.

\section{Diffeomorphism invariant gauge theories}
\label{sec:formulation}

\subsection{Gravity as a gauge theory}

In the pure connection formulation gravity becomes the most general diffeomorphism invariant gauge theory. In the case of a purely gravitational theory\footnote{One can also consider unified Yang-Mills-gravity theories of the same sort, see \cite{TorresGomez:2010cd}.} the gauge group is (complexified) ${\rm SU}(2)\sim{\rm SO}(3)$. The action is a functional of an ${\rm SU}(2)$ connection $A^i, i=1,2,3$ on a spacetime manifold $M$. Let $F^i=dA^i+(1/2)\epsilon^{ijk}A^j\wedge A^k$ be the curvature of $A^i$. The action is given by the following gauge and diffeomorphism invariant functional of the connection:
\be\label{action}
S[A]=(1/\im) \int_M f(F^i\wedge F^j).
\ee
Here $\im=\sqrt{-1}$ is a factor introduced for future convenience, and $f$ is a function with properties to be spelled out below. 

We shall refer to $f$ as the defining function of our theory. It is a holomorphic, homogeneous of degree one and gauge invariant function of its matrix (and 4-form) valued argument. Thus, let $X^{ij}\in {\mathfrak su}(2)\otimes_S{\mathfrak su}(2)$ be a matrix valued in the second symmetric power of the Lie algebra. The gauge group ${\rm SU}(2)\sim{\rm SO}(3)$ acts in the space of such matrices via $X\to g X g^T$, where $T$ is the operation of the transpose. We first consider scalar valued functions $f: {\mathfrak su}(2)\otimes_S{\mathfrak su}(2) \to \C$ that are holomorphic, gauge-invariant $f(g X g^T)=f(X)$ and homogeneous of degree one $f(\alpha X)=\alpha f(X)$. A convenient for practical computations parameterization of such functions is as follows. Consider the following 3 ${\rm SU}(2)$ invariants of $X^{ij}$:
\be
{\rm Tr}(X), \qquad {\rm Tr}(X^2), \qquad {\rm Tr}(X^3),
\ee
where the traces (and powers of $X$) are computed using the Killing metric $\delta^{ij}$ on the Lie algebra. When ${\rm Tr}(X)\not=0$ we can parameterise the defining function $f$ as follows:
\be\label{f}
f(X) = {\rm Tr}(X)\, \chi\left( \frac{{\rm Tr}(X^2)}{({\rm Tr}(X))^2}, \frac{{\rm Tr}(X^3)}{({\rm Tr}(X))^3}\right),
\ee
where $\chi$ is now an arbitrary holomorphic function of its two arguments.

Given $f$ with the properties as spelled out above, e.g. one parameterised as in (\ref{f}), it can be seen that this function can be applied to a matrix valued 4-form, with the result being a 4-form. Indeed, consider $F^i\wedge F^j$, which is a ${\mathfrak su}(2)\otimes_S{\mathfrak su}(2)$ valued 4-form. Choose a reference volume form on $M$ (we assume that $M$ is orientable), and denote it by $(\rm vol)$. Of course, $(\rm vol)$ is only defined modulo the multiplication by a nowhere zero function. Using this reference volume form we can write $F^i\wedge F^j = X^{ij} ({\rm vol})$, where $X^{ij}$ is again defined only modulo rescalings. We can now use the homogeneity of $f$ to write
\be
f(F^i\wedge F^j)=({\rm vol}) f(X).
\ee
It is moreover clear that the result on the right-hand-side does not depend on which reference volume form is used in this argument. This is again due to the homogeneity of $f$. This shows that the integrand in (\ref{action}) is a well-defined 4-form that can be integrated to obtain the action. This finishes the formulation of our theory.

We note that, as formulated, there are no dimensionful parameters in our theory. Indeed, we assume the connection field $A^i$ to have the usual mass dimension one, so that the curvature has the mass dimension two, and the matrix of the wedge products $[X]=4$. The defining function $f$ is essentially the function $\chi$ of ratios of powers of $X^{ij}$ that are dimensionless, and so does not contain any dimensionful parameters (but contains an infinite number of dimensionless "coupling constants", once expanded appropriately). Thus, due to the homogeneity of $f$, its mass dimension is the same as that of $X$ (in the parameterization (\ref{f}) the mass dimension is carried by the first term ${\rm Tr}(X)$, while the function $\chi$ is dimensionless). The function $f$ can then be integrated to produce a dimensionless action (as usual we work in the units $c=\hbar=1$). As we shall see, the fact that there are no dimensionful coupling constants in our theory has profound implications for the structure of its perturbation theory.

Classically (\ref{action}) is a theory that can be shown, see e.g. \cite{Krasnov:2007cq}, to propagate two (complex for the time being, reality conditions will be discussed below) degrees of freedom. We will see a version of the argument that leads to this conclusion below when we consider the perturbation theory. One can also show that theory (\ref{action}) is a gravity theory, in spite of the fact that no metric is present anywhere. Thus, it can be reformulated explicitly as a theory of metrics via a sequence of transformations. The main idea is to note that declaring the 3 two-forms $F^i$ to span the space of (anti-) self-dual two-forms determines a conformal metric on $M$ whenever the matrix $X^{ij}$ defined from the wedge product of curvatures is non-degenerate. One can then rewrite the theory (\ref{action}) explicitly as the theory of this metric, see \cite{Krasnov:2009ik} for details. We also note that the usual general relativity (with or without the cosmological constant) can be rewritten in this language, see below. However, in this paper, we shall not need this relation to metric theories. Our plan is to study (\ref{action}) as is. We shall set the stage for its perturbative quantization and a study of its renormalization. The main justification for this undertaking is that a whole class of gravity theories (for varying defining functions $f$) can be treated in one go. Moreover, our theories are theories of a connection, and we can hope to use the expertise that was accumulated in quantum field theory for dealing with quantum gauge theories. 

\subsection{GR with the cosmological constant}

Before we proceed with our analysis of theories (\ref{action}), we would like to state the action principle that reformulates the usual GR (with the cosmological constant) in this language. Consider the following action principle:
\be\label{sec-GR-action}
S_{\rm{GR}}[A] = \frac{1}{16\pi\im G\Lambda} \int \left( {\rm Tr}\sqrt{F^i\wedge F^j} \right)^2.
\ee
Here $ {\rm Tr}\sqrt{F^i\wedge F^j} $ is the trace of a matrix square root of the matrix $F^i\wedge F^j$. It is clear that the above action is of the general form (\ref{action}), for the scalar function that is used in the action (\ref{sec-GR-action}) is gauge-invariant and homogeneous of degree one, as required. The quantities $G$, $\Lambda$ in the denominator in front of the action are the usual Newton's constant and the cosmological constant respectively. Note that these appear in the action only in the dimensionless combination $G\Lambda$. For the currently accepted value of $\Lambda$ the value of $G\Lambda$ is exceedingly small $G\Lambda\sim M_\Lambda^2/M_p^2\sim 10^{-120}$. Thus, the value of the dimensionless parameter in front of GR action is very large. Below we shall see what kind of implication this has for the gravity perturbation theory. For the convenience of the reader the action (\ref{sec-GR-action}) is derived in the Appendix. 

\subsection{Topological action}

Another prominent member of the class of theories (\ref{action}) is the topological theory:
\be\label{top-action}
S_{\rm top}[A]= \frac{1}{\im \kappa} \int {\rm Tr} (F^i\wedge F^j),
\ee 
where $\kappa$ is some numerical parameter. It is not hard to see that the Lagrangian is a total derivative, and so the action describes a theory without propagating DOF --- a topological theory. Being topological, this theory is certainly a fixed point of the (sought) renormalization group flow in the space of theories (\ref{action}). Conjecturally, the renormalization group flow takes one from (\ref{sec-GR-action}) at low energies to (\ref{top-action}) at high energies. 

\subsection{A convex functional}

We would also like to give an example of an action with a convex (near the point $X^{ij}\sim \delta^{ij}$) defining function. Let us consider the following theory:
\be\label{joel-action}
S[A] = \frac{1}{\im\tilde{\kappa}} \int \frac{{\rm Tr}(F^i\wedge F^j)^2} {{\rm Tr}(F^i\wedge F^j)},
\ee
where $\tilde{\kappa}$ is a dimensionless parameter. The downward gradient flow for this functional of the connection has been studied in \cite{Joel}. The renormalization group flow for the class of theories (\ref{action}) should in particular explain why at low energies one flows to a concave functional (\ref{sec-GR-action}) instead of a convex functional such as (\ref{joel-action}).

\subsection{First variation and field equations}

The first variation of the action (\ref{action}) gives us the field equations. To write these down, let us give a parameterization of the matrix $X^{ij}$ useful for practical computations. Thus, let $\tilde{\epsilon}^{\mu\nu\rho\sigma}$ be a completely anti-symmetric rank 4 vector, which is a density of weight one (as is indicated by the tilde over its symbol). This object exists on any orientable manifold and does not need a metric for its definition. Consider:
\be\label{X}
\tilde{X}^{ij}:= \frac{1}{4} \tilde{\epsilon}^{\mu\nu\rho\sigma} F^i_{\mu\nu} F^j_{\rho\sigma},
\ee
where as before $F^i_{\mu\nu}$ is the curvature two-form, with its spacetime indices now indicated explicitly. The quantity $\tilde{X}^{ij}$ is a ${\mathfrak su}(2)\otimes_S{\mathfrak su}(2)$ valued matrix, and a density of weight one. One takes the defining function $f$ to be a function of $\tilde{X}^{ij}$ given by the same expression as in (\ref{f}). With convention $dx^\mu \wedge dx^\nu\wedge dx^\rho \wedge dx^\sigma=\tilde{\epsilon}^{\mu\nu\rho\sigma} d^4x$ we can write the action (\ref{action}) as
\be
S[A] = (1/\im) \int_M d^4x\, f(\tilde{X}^{ij}) \, .
\ee
The first variation of the action can now be easily computed and reads:
\be\label{first-var}
\delta S[A]=(1/\im) \int_M d^4x\, \frac{\partial f}{\partial \tilde{X}^{ij}} \frac{1}{2} \tilde{\epsilon}^{\mu\nu\rho\sigma} F^i_{\mu\nu} D_{A\,\rho} \delta A^j_\sigma \, .
\ee
Integrating by parts, we see that the field equations for (\ref{action}) can be written as
\be\label{feqs}
D_A B^i =0,
\ee
where we have used the form notations again, and the two-form $B^i$ is defined via
\be\label{B}
B^i:= \frac{\partial f}{\partial \tilde{X}^{ij}} F^j.
\ee
We note that the matrix of first derivatives that appear on the right-hand-side of this expression is a symmetric matrix, and has density weight zero (as the ratio of the density weight one function $f(\tilde{X})$ and the density weight one quantity $\tilde{X}$). Thus, (\ref{B}) is a well-defined two-form. 

For example, in the case of GR we get:
\be
B^i_{\rm GR} = \frac{{\rm Tr}\sqrt{X}}{16\pi G\Lambda} ((\sqrt{X})^{-1})^{ij} F^j.
\ee
Then, using the definition of $X^{ij}\sim F^i\wedge F^j$ we can easily see that in the case of GR
\be
B^i_{\rm GR}\wedge B^j_{\rm GR}\sim \delta^{ij},
\ee
which is the usual "metricity" equation of Plebanski formulation of GR \cite{Plebanski:1977zz}. Another example is that of the topological theory (\ref{top-action}). In this case the $B$-field is given by:
\be
B_{\rm top}^i = \frac{1}{\kappa} F^i,
\ee
and the field equation (\ref{feqs}) is satisfied automatically as a consequence of the Bianchi identity $D_A F=0$.

\subsection{Symplectic structure}

The computation of the first variation in the previous subsection also gives us the symplectic structure of the theory. Thus, the phase space of the theory is the space of all solutions of (\ref{feqs}), and the symplectic structure can be obtained by considering the boundary term that was neglected in passing from (\ref{first-var}) to (\ref{feqs}). The integral of the boundary term gives rise to an integral over the spatial slice $\Sigma$ of the following quantity
\be
\Theta := \frac{1}{2\im} \int_\Sigma B^i \wedge \delta A^i,
\ee
where $B^i$ is as in (\ref{B}). This is a one-form on the phase space of the theory. Its exterior derivative produces the symplectic two-form. We see that the significance of the quantity $B^i$ defined by (\ref{B}) is that its spatial projection plays the role of the momentum canonically conjugate to the spatial projection of the connection $A^i$. We emphasise that in the present "pure gauge" formulation, the two-form $B^i$ is not independent and is a function of the connection field. A formulation that "integrates in" the two-form field as an independent variable is possible, and has been studied in previous works by the author, but will not be considered here. 

\subsection{Gauge invariance}

Let us now verify by an explicit computation that our theory is invariant under diffeomorphisms as well as ${\rm SO}(3,\C)$ rotations. This is of course expected, because the action was constructed in the way that these invariances should hold. However, an explicit verification of this fact will allow us to establish some identities for the latter. The gauge transformations act on the connection field as follows
\be
\delta_\xi A^i_\mu = \xi^\alpha F_{\mu\alpha}^i, \qquad \delta_\phi A^i_\mu = D_{A\, \mu}\phi^i.
\ee
The first of these transformations can be seen to be a diffeomorphism corrected by a gauge transformation, while the second one is the usual gauge rotation with the parameter $\phi^i$.

It is not too difficult to prove the invariance of our action (\ref{action}) under these transformations. Let us first consider the diffeomorphisms. The variation of the action (\ref{first-var}) becomes proportional to
\be\label{0-diff-inv-1}
 \int_M d^4x\, \frac{\partial f}{\partial \tilde{X}^{ij}} \tilde{\epsilon}^{\mu\nu\rho\sigma} F^i_{\mu\nu} D_{A\,\rho} \xi^\alpha F^j_{\sigma\alpha}.
 \ee
 We now need some identities. First we note that one can write the Bianchi identity $D_A F^i=0$ as
 \be\label{bianchi}
 D_{A\,[\mu} F^i_{\nu]\rho}=-\frac{1}{2}D_{A\,\rho} F_{\mu\nu}^i.
 \ee
 Another identity that we need is 
 \be\label{FF-ident}
 \tilde{\epsilon}^{\mu\nu\rho\sigma} F^{(i}_{\mu\nu}  F^{j)}_{\sigma\alpha}=-\frac{1}{4} \delta^\rho_\alpha \tilde{\epsilon}^{\mu\nu\gamma\delta} F^i_{\mu\nu}  F^j_{\gamma\delta}=-\delta^\rho_\alpha \tilde{X}^{ij},
 \ee
 where $\delta^\rho_\alpha$ is the Kronecker delta. Note that the symmetrisation is taken on the left hand-side. The above two identities, as well as the definition (\ref{X}) of the matrix $\tilde{X}^{ij}$, allow us to rewrite (\ref{0-diff-inv-1}) as
 \be\label{0-diff-inv-4}
  -\int_M d^4x\, \frac{\partial f}{\partial \tilde{X}^{ij}} \left( \tilde{X}^{ij} \partial_\alpha \xi^\alpha +
  \frac{1}{2} \tilde{\epsilon}^{\mu\nu\rho\sigma} F^i_{\mu\nu} \xi^\alpha D_{A\,\alpha} F_{\rho\sigma}^j\right)=
  -\int_M d^4x\, \frac{\partial f}{\partial \tilde{X}^{ij}} D_{A\,\alpha} ( \xi^\alpha \tilde{X}^{ij}) .
  \ee
  Integrating by parts, this becomes equal to
  \be\label{0-diff-inv-2}
  \int_M d^4x\, \xi^\alpha \tilde{X}^{ij} D_{A\,\alpha} \frac{\partial f}{\partial \tilde{X}^{ij}} .
  \ee
  We should now see that the integrand here is zero. This follows from the homogeneity of the function $f$. Indeed, we have  
  \be\label{f-hom}
  \tilde{X}^{ij}  \frac{\partial f}{\partial \tilde{X}^{ij}} =f
  \ee
  from the fact that $f$ is a homogeneous function of degree one. Let us now apply the operator of partial derivative $\partial_\mu$ to both sides of this equation. We get
  \be
 ( \partial_\mu \tilde{X}^{ij} ) \frac{\partial f}{\partial \tilde{X}^{ij}} + \tilde{X}^{ij} \partial_\mu \frac{\partial f}{\partial \tilde{X}^{ij}} = \partial_\mu f = \frac{\partial f}{\partial \tilde{X}^{ij}} \partial_\mu \tilde{X}^{ij}.
 \ee
 Comparing the two sides we see that
 \be\label{X-df}
 \tilde{X}^{ij} \partial_\mu \frac{\partial f}{\partial \tilde{X}^{ij}} =0,
 \ee
 which is almost the integrand in (\ref{0-diff-inv-2}), except for the fact that we have the covariant derivative in (\ref{0-diff-inv-2}). Let us now consider the difference between the covariant and the usual derivatives. We have
 \be\label{0-diff-inv-3}
  \tilde{X}^{ij}  (D_{A\,\mu} - \partial_\mu ) \frac{\partial f}{\partial \tilde{X}^{ij}}=
 2  \tilde{X}^{ij} \epsilon^{ikl} A_\mu^k \frac{\partial f}{\partial \tilde{X}^{lj}}.
   \ee
 The quantity here is zero in view of the gauge invariance of the function $f$. Indeed, under infinitesimal gauge transformations an ${\mathfrak su}(2)\otimes_S{\mathfrak su}(2)$-valued matrix $\tilde{X}^{ij}$ transforms as
 \be
 \delta_\phi \tilde{X}^{ij} = \epsilon^{ikl} \phi^k \tilde{X}^{lj} + \epsilon^{jkl} \phi^k \tilde{X}^{il}.
 \ee
 Then the statement that $f$ is an ${\rm SO}(3,\C)$ invariant function becomes
  \be\label{X-comm-df}
 \epsilon^{ikl} \tilde{X}^{kj}  \frac{\partial f}{\partial \tilde{X}^{lj}}  = 0,
 \ee
which can be expressed in words by saying that the commutator of the matrix $\tilde{X}^{ij}$ with the matrix $\partial f/\partial \tilde{X}^{ij}$ of the first derivatives of the defining function is zero.

The identity (\ref{X-comm-df}) immediately implies that the difference of the derivatives in (\ref{0-diff-inv-3}) is zero and thus 
 \be\label{X-Df}
 \tilde{X}^{ij} D_{A\,\mu} \frac{\partial f}{\partial \tilde{X}^{ij}} =0,
 \ee
 which proves the invariance of the action (\ref{action}) under diffeomorphisms.

 Let us now prove the invariance of (\ref{action}) under the gauge rotations. The variation of the action in this case becomes proportional to
 \be
  \int_M d^4x\, \frac{\partial f}{\partial \tilde{X}^{ij}} \tilde{\epsilon}^{\mu\nu\rho\sigma} F^i_{\mu\nu} D_{A\,\rho} D_{A\,\sigma} \phi^j.
 \ee
 Expressing the commutator of the covariant derivatives as the commutator with the curvature, and recalling the definition (\ref{X}) of the matrix $\tilde{X}^{ij}$ we get
 \be
 4 \int_M d^4x\, \frac{\partial f}{\partial \tilde{X}^{ij}} \epsilon^{jkl} \tilde{X}^{ik} \phi^l,
 \ee
 which is zero in view of (\ref{X-comm-df}). This proves the invariance of the action (\ref{action}) under the ${\rm SO}(3,\C)$ rotations.

\subsection{Second variation}

We can now compute the second variation, in preparation for the next section treatment. We have
\be\label{sec-var}
\delta^2 S[A] = (1/\im) \int_M d^4x\, \left( \frac{\partial^2 f}{\partial \tilde{X}^{ij}\partial \tilde{X}^{kl}} \delta \tilde{X}^{ij} \delta \tilde{X}^{kl}
+\frac{\partial f}{\partial \tilde{X}^{ij}} \delta^2 \tilde{X}^{ij} \right).
\ee
Here the first variation of $\tilde{X}^{ij}$ was already computed above and reads
\be\label{X-1}
\delta \tilde{X}^{ij} =\frac{1}{2} \tilde{\epsilon}^{\mu\nu\rho\sigma} F^{(i}_{\mu\nu} D_{A\,\rho} \delta A^{j)}_\sigma\, .
\ee
The second variation reads
\be\label{X-2}
\delta^2 \tilde{X}^{ij} =\frac{1}{2} \tilde{\epsilon}^{\mu\nu\rho\sigma} D_{A\,\mu} \delta A^i_\nu D_{A\,\rho} \delta A^j_\sigma +
 \frac{1}{2} \tilde{\epsilon}^{\mu\nu\rho\sigma} F^{(i}_{\mu\nu} \epsilon^{j)kl} \delta A^k_\rho \delta A^l_\sigma \, .
 \ee

\section{Constant curvature background}
\label{sec:backgr}

In this and the next section, to get a better feel for our theory and also to prepare for its quantization, we consider the action (\ref{action}) expanded around a specific background connection $A^i$. 

\subsection{Second order action around a general background}

We now write our connection as the background $A^i$ plus a fluctuation $\A^i$, and obtain the part of the action quadratic in $\A^i$ directly from (\ref{sec-var}). Thus, we divide the second variation by 2, replace $\delta A^i_\mu$ by $\A_\mu^i$, and get the following Lagrangian
 \be\label{act-2}
 (8\im) {\cal L}_\A =   \frac{\partial^2 f}{\partial \tilde{X}^{ij}\partial \tilde{X}^{kl}} (\tilde{\epsilon}^{\mu\nu\rho\sigma} F^i_{\mu\nu} D_{A\,\rho} \A^j_\sigma) (\tilde{\epsilon}^{\alpha\beta\gamma\delta} F^k_{\alpha\beta} D_{A\,\gamma} \A^l_\delta)
 \\ \nonumber
+  2\frac{\partial f}{\partial \tilde{X}^{ij}} \tilde{\epsilon}^{\mu\nu\rho\sigma} \left( D_{A\,\mu} \A^i_\nu D_{A\,\rho} \A^j_\sigma 
+ F^{i}_{\mu\nu} \epsilon^{jkl} \A^k_\rho \A^l_\sigma \right) \, .
 \ee
In the following works this action will be used for a background field method one-loop computation, but here we specialise to a particular background.

\subsection{The background}

The background that we take is a constant curvature one and can be defined as follows. As we have already mentioned, any connection defines a (conformal) metric, obtained by requiring the triple of curvature two-forms $F^i$ to be (anti-)self-dual. Because  of this, in practice, to specify a background it is easier to start with the corresponding metric, and then construct the connection, so that the triple of curvature two-forms for this connection is (anti-)self-dual with respect to the metric one started from. This is the procedure we follow. 

So, we first describe the corresponding metric, and then use it to construct the background connection in question. Thus, let $ds^2$ be the interval for a constant curvature metric in 4 spacetime dimensions (de Sitter space). For our purposes it is convenient to describe it using the flat slicing so that the metric reads:
 \be\label{metric-backgr}
 ds^2 = a^2(\eta)( - d\eta^2 + \sum_{i=1}^3 (dx^i)^2),
 \ee
 where $\eta$ is the conformal time and $x^i$ are the spatial coordinates. For the de Sitter metric the function $a^2(\eta)$ is a specific one, see below. The tetrad $\theta^I, I=0,1,2,3$ associated to the above metric reads:
 \be\label{tetrad}
 \theta^0 = a d\eta, \qquad \theta^i = a dx^i,
 \ee
 so that $ds^2=\theta^I\otimes \theta^J\eta_{IJ}$, where $\eta_{IJ}={\rm diag}(-1,1,1,1)$. As is known to anyone with experience with the Plebanski formulation of General Relativity \cite{Plebanski:1977zz}, it is very convenient to define the following set of objects (two-forms):
\be\label{Sigma}
\Sigma^i := \im \theta^0 \wedge \theta^i - \frac{1}{2} \epsilon^{ijk}\theta^j \wedge \theta^k,
\ee
where, as before $i=1,2,3$. Explicitly, for the metric (\ref{metric-backgr}) we have:
\be\label{Sigma-backgr}
\Sigma^i = a^2 \left( \im d\eta\wedge dx^i -\frac{1}{2} \epsilon^{ijk} dx^j \wedge dx^k\right).
\ee
As is not hard to check, in general the two-forms (\ref{Sigma}) are anti-self-dual 
\be\label{asd}
\frac{\im}{2}\epsilon_{\mu\nu}{}^{\rho\sigma} \Sigma^i_{\rho\sigma}=\Sigma^i_{\mu\nu}
\ee
with respect to the Hodge star operation on two-forms defined by the metric $ds^2=\theta^I\otimes \theta^J\eta_{IJ}$. Here the object $\epsilon_{\mu\nu}{}^{\rho\sigma}$ is obtained from the volume form $\epsilon_{\mu\nu\rho\sigma}$ by raising two of its indices using the metric, and in our conventions $\epsilon^{0123}=+1$. Thus, (\ref{Sigma-backgr}) are anti-self-dual with respect to the metric (\ref{metric-backgr}). 

Let us now introduce our background connection. It is ${\rm SU}(2)$ connection $A_0^i$ such that the covariant exterior derivative of $\Sigma^i$ given by (\ref{Sigma-backgr}) with respect to $A_0^i$ is zero. In other words:
\be
0=D_{A_0} \Sigma^i =d\Sigma^i + \epsilon^{ijk} A_0^j \wedge \Sigma^k.
\ee
It is not hard to solve this equation for $A_0^i$ explicitly. We get
\be\label{conn-backgr}
A_0^i = \im \cH dx^i,
\ee
where
\be
\cH=\frac{a'}{a},
\ee
and the prime denotes the derivative with respect to the conformal time. It is not hard to show that the connection (\ref{conn-backgr}) is just the (anti-) self-dual part of the spin connection compatible with the tetrad (\ref{tetrad}).

We have not yet used (imposed) the condition that the background (\ref{metric-backgr}) is constant curvature. This condition can be written as
\be\label{curv-0}
F^i(A_0) = M_0^2 \Sigma^i,
\ee
where we have introduced a dimensionful parameter $M_0$, with dimensions of mass. The equation (\ref{curv-0}) states that the curvature of $A^i$ is a constant $M_0^2$. For the connection (\ref{conn-backgr}) this gives two equations:
\be
\cH'=\cH^2=a^2 M_0^2,
\ee
where the second equality is the familiar Friedman equation. Its solution can be written as $a=(M_0 (\eta_{max}-\eta))^{-1}$, where $\eta_{max}$ is an integration constant. This means that the physical time $t$ (obtained from $a d\eta=dt$) is determined by the relation $(M_0 (\eta_{max}-\eta))^{-1}=e^{M_0 t}$, where a convenient choice of the integration constant was made. Thus we have $a=e^{M_0 t}$ and therefore an exponentially expanding Universe. The discussed constant curvature metric (de Sitter space) is of course a solution of Einstein's theory. The quantity $M_0$ is then related to the cosmological constant $\Lambda$ via $M_0^2=\Lambda/3$. In the case of GR the cosmological constant is a parameter of the theory. In the case of our theories, however, there is no similar parameter in the Lagrangian, so we will not in general be able to identify $M_0$ with any $\Lambda$, as there is no such parameter in the theory. However, we shall see below that a certain analog of $\Lambda$ can be defined for any of our theories by evaluating the action on the background (\ref{conn-backgr}). 

We thus take the constant curvature connection 
\be\label{A0}
A_0^i= \im M_0 a dx^i.
\ee
 as the background for the perturbative expansion of (\ref{action}). Note that, as far as the background is concerned, the flat limit $M_0\to 0$ can be taken without any difficulty. In this limit $a\to 1$ and $A_0^i\to 0$. 
 
 \subsection{Action evaluated on the backgorund}
 
In GR the gravitational action evaluated on the de Sitter metric is proportional to the volume of the Universe. Indeed,  the Einstein-Hilbert action for the signature $(-,+,+,+)$ reads
 \be\label{EH}
 S_{\rm EH}[g]=-\frac{1}{16\pi G} \int (R-2\Lambda)\sqrt{-g}\, d^4x.
 \ee
 On a constant curvature background (in 4 dimensions) $R=4\Lambda$ and we get
 \be\label{SEH0}
 S^0_{\rm EH}= -\frac{\Lambda}{8\pi G} \int \sqrt{-g}\, d^4x.
 \ee

Let us see what our action evaluated on (\ref{A0}) gives us. Using 
 \be
 \tilde{\epsilon}^{\mu\nu\rho\sigma} \theta_\mu^0 \theta_\nu^i \theta_\rho^j\theta_\sigma^k = \sqrt{-g} \epsilon^{ijk},
 \ee
 where $\sqrt{-g}$ is the square root of the determinant of the metric $ds^2=\theta^I\otimes \theta^J\eta_{IJ}$, we easily get
 \be
 \Sigma^i\wedge \Sigma^j = -2\im \sqrt{-g} \, \delta^{ij} d^4x\, ,
 \ee
 where $\Sigma^i$ are the anti-self-dual forms (\ref{Sigma}). Thus, the matrix $\tilde{X}^{ij}$ at the background is equal to
 \be\label{X0}
 \tilde{X}^{ij}_0 = -2\im M_0^4 \sqrt{-g} \, \delta^{ij}\, ,
 \ee
 i.e., is proportional to the identity matrix. Thus, the value of the action (\ref{action}) on the background is
 \be\label{backgr-action}
 S[A_0]=-2M_0^4 f_0 \int \sqrt{-g} \, d^4x,
 \ee
 where $f_0:=f(\delta)$ is the value of the defining function at the identity matrix $X^{ij}=\delta^{ij}$. 
 
Thus, for the function $f$ that corresponds to GR we expect (\ref{SEH0}) to be equal to (\ref{backgr-action}) and thus
 \be\label{M0-Lambda}
 2M_0^4 f_0 = \frac{\Lambda}{8\pi G}.
 \ee
 As is shown in the Appendix for the defining function of GR $f_0= 9/(16\pi G\Lambda)$, and so the relation (\ref{M0-Lambda}) holds. For diffeomorphism invariant gauge theories different from GR we have neither $G$ nor $\Lambda$ parameters. The only parameters of the theory are those (dimensionless) parameters arising by expanding the defining function. The sole dimensionful parameter comes in when the background curvature parameter $M_0$ is chosen. The relation (\ref{M0-Lambda}) then shows that we have a natural analog of the ratio $\Lambda/G$ present in our theory, and given by the product of the defining function evaluated at the identity matrix times $M_0^4$. However, there is yet no natural way to define either $\Lambda$ or $G$. Indeed, we could choose to expand a theory with given $f$ around background with any value of $M_0$, so there is no reason for the identification $M_0^2=\Lambda/3$ as in GR. Similarly, without analyzing how our gravitons interact it is impossible to determine any analog of the Newton's constant for our theory. We note, however, that since for GR we have  $f_0\sim 10^{120}$, we should expect that for the defining functions of interest the value of the defining function on the identity matrix is extremely large. This knowledge will be of help when we analyze the graviton self-interactions.  
  
  \subsection{Linearized action}
 
 We first check that the constant curvature background (\ref{A0}) is a solution of (\ref{feqs}) and then evaluate the second variation of the action (\ref{sec-var}) at the background.
 
 The derivatives of (\ref{f}) at the identity matrix are easily computed. Let us first write down the general expression for the first derivative. We omit the tilde from $X$ for brevity (we can always pull out the density weight factor from the function $f$ using the homogeneity). We have
 \be\label{der-1}
 \frac{\partial f}{\partial X^{ij}} = \delta^{ij} \chi(X) + {\rm Tr}(X)\chi'_1(X) \left( \frac{2X^{ij}}{({\rm Tr}(X))^2} - \frac{2{\rm Tr}(X^2)}{({\rm Tr}(X))^3} \delta^{ij} \right) 
 \\ \nonumber
 +  {\rm Tr}(X)\chi'_2(X) \left( \frac{3(X^2)^{ij}}{({\rm Tr}(X))^3} - \frac{3{\rm Tr}(X^3)}{({\rm Tr}(X))^4} \delta^{ij} \right),
 \ee
 where $\chi_{1,2}'(X)$ are the derivatives of the function $\chi$ with respect to the first and second arguments, evaluated at $X$. It is easy to check that for $X^{ij}_0\sim \delta^{ij}$ the second and third terms on the right are zero, and we have:
 \be\label{fp}
   \frac{\partial f}{\partial X^{ij}}\Big|_{X_0} = \delta^{ij} \chi(X_0) = \frac{f_0}{3} \delta^{ij}.
   \ee
 We note that this is $M_0$ independent. We remind the reader that the background value $X_0$ of matrix $X$ is given by (\ref{X0}) above.

Let us now compute the matrix of second derivatives of the defining function. Since the expressions in brackets in (\ref{der-1}) become zero when evaluated on $X_0$, the only way to get a non-zero result in the second derivative is to act by a derivative on these expressions. We get
\be\label{fpp}
\frac{\partial^2 f}{\partial X^{ij}\partial X^{kl}}\Big|_{X_0} = \frac{2(\chi_1'(X_0)+\chi_2'(X_0))}{{\rm Tr}(X_0)} P^{ij|kl} \, ,
\ee 
where
\be
 P^{ij|kl}:= I^{ij|kl} -\frac{1}{3}\delta^{ij}\delta^{kl}, \qquad I^{ij|kl} := \frac{1}{2} \left(\delta^{ik}\delta^{jl}+\delta^{il}\delta^{jk}\right).
\ee
We have introduced a special notation $P^{ij|kl}$ for the matrix that appeared in (\ref{fpp}), as this is just the projector on the symmetric traceless part $P^{ij|kl} \delta_{ij} = 0$, and similarly for the contraction with $\delta_{kl}$.

Having evaluated the derivatives of the defining function at the background, we are ready to specialise (\ref{act-2}) for our constant curvature background (\ref{A0}). However, let us first check that our chosen background is indeed a solution of field equations (\ref{feqs}). With the quantity $\chi(X_0)$ being a constant, the background $B^i_0\sim F^i(A_0)$, and thus the field equations (\ref{feqs}) are satisfied (in view of the Bianchi identity). 

Let us now consider the second term in (\ref{sec-var}). Since the matrix of the first derivatives is proportional to the identity matrix (\ref{fp}) with a constant proportionality coefficient we need to consider the integral of $\delta_{ij} \delta^2 \tilde{X}^{ij}$ over the manifold. Let us see that this is a total derivative. We have:
\be\label{comp-sec-var-1}
\int_M d^4x \, \delta_{ij} \delta^2 \tilde{X}^{ij} = \frac{1}{2} \int_M \left( D_A \delta A^i \wedge D_A \delta A^i + 
F^i(A) \wedge \epsilon^{ijk}  \delta A^j \wedge \delta A^k \right),
\ee
where we wrote everything in terms of forms (our form convention is $F=(1/2)F_{\mu\nu} dx^\mu\wedge dx^\nu$). Integrating by parts in the first term (and neglecting the total derivative term), the first term becomes
\be
 \frac{1}{2} \int_M \delta A^i \wedge D_A D_A \delta A^i =  \frac{1}{2} \int_M \delta A^i \wedge \epsilon^{ijk} F^j(A) \wedge \delta A^k,
 \ee
 which is minus the second term in (\ref{comp-sec-var-1}), and so (\ref{comp-sec-var-1}) is a total derivative. 
 
 We therefore only need to consider the first term in (\ref{sec-var}). Let us write this directly in terms of the two-forms $\Sigma^i$ by substituting the expression (\ref{curv-0}) for the background curvature. Using the anti-self-duality (\ref{asd}) of $\Sigma^i$ we have the following compact expression for the second variation
 \be\label{lin-act-1}
 \delta^2 S \Big|_{A_0}= - g_0 \int_M d^4x \sqrt{-g} \, P^{ij|kl} (\Sigma^{i\,\mu\nu} D_{A_0\, \mu} \delta A^j_\nu) (\Sigma^{k\,\rho\sigma} D_{A_0\, \rho} \delta A^l_\sigma),
 \ee
 where we have introduced a notation
 \be\label{g}
 g_0:= \frac{\chi_1'(X_0)+\chi_2'(X_0)}{3}.
 \ee
 Note that the factors of $M_0$ have cancelled from this result. The combination (\ref{g}) of the first derivatives of the defining function plays an important role below. Thus, we shall see that the constant determining the strength of self-interactions of our gravitons will be built from $M_0$ and $g_0$. We note that for the case of GR the quantity $g_0$ is of the same order as $f_0$ (see Appendix), and thus is very large. 
  
 \subsection{High energy limit}
 
For applications in quantum gravity one is mostly interested in the UV behaviour of the theory. Thus, we are interested in its behaviour at energies $E\gg M_0$. In this case we can neglect the fact that the background is curved, and consider an effective theory in Minkowski space. As we shall see in this subsection, this certainly works at the limearized level. At the level of interactions we shall face some puzzles (related to the fact that the GR action blows up in the limit $\Lambda\to 0$), to be discussed below. 
 
At energies $E\gg M_0$ the terms in the covariant derivative containing the usual derivative become much larger than the terms containing the background connection (the latter being of the order $M_0$). Thus, in the high energy limit we can replace the covariant derivatives with the ordinary ones, and neglect the fact that the background is curved. However, the field $\delta A^i$ in the linearized action (\ref{lin-act-1}) is not canonically normalized, as there is a numerical constant $g_0$ in front of the action. Absorbing this constant into the linearized fields by rescaling we obtain the following action
 \be\label{lin-action}
 S_{\rm lin}[a] = -\frac{1}{2} \int_M d^4x \, P^{ij|kl} (\Sigma^{i\,\mu\nu} \partial_\mu a^j_\nu) (\Sigma^{k\,\rho\sigma}\partial_\rho a^l_\sigma),
 \ee
 where the (rescaled) linearized field is now called $a^i_\mu=\sqrt{g_0}(\delta A_\mu^i)$, and we have divided the second variation of the action by 2 to get the correct linearized action. The two-forms $\Sigma^i_{\mu\nu}$ are now those corresponding to the Minkowski spacetime
 \be
 \Sigma^i = \im dt\wedge dx^i - \frac{1}{2} \epsilon^{ijk} dx^j \wedge dx^k.
 \ee
 Thus, in the high energy limit one effective works in the Minkowski background, and the connection perturbation has been rescaled to have a canonically normalized kinetic term. The operation of absorbing $g_0$ into the connection field is not that innocuous, as $g_0$ blows up in the limit $\Lambda\to 0$. But if one is not taking this limit, just considers the connection perturbations changing on scales much smaller than the scale of the curvature, then it is natural to absorb the (very large) quantity $g_0$ into the connection field to make it canonically normalized. We shall now study the action (\ref{lin-action}) in some detail, to see that it does describe the usual Minkowski spacetime gravitons. After this we turn to interactions. 
 
 A quick note about dimensions of all the fields. As we have already mentioned, we take the connection to have the mass dimension one, as is appropriate for a field that can be combined into a derivative operator. Then the curvature has mass dimension two, the matrix $X^{ij}$ has mass dimension 4, the matrix of first derivatives of the defining function is dimensionless, and the matrix of second derivatives has dimension minus 4. The two-forms $\Sigma^i$ that are constructed from the dimensionless metric are dimensionless. The constant $g_0$ introduced in (\ref{g}) is a sum of derivatives of a function of dimensionless arguments, and thus is dimensionless. Overall, we see that the mass dimension of the integrand in (\ref{lin-action}) is 4, as needed. 
 
 \subsection{Symmetries}
 
 We have started from a diffeomorphism invariant action (\ref{action}) and linearized it around the constant curvature (and then zero curvature) background. We should check that the linearized action that we have obtained is still diffeomorphism invariant. As before, the diffeomorphisms can be lifted to the ${\rm SU}(2)$ bundle as follows:
 \be
 \delta_\eta A^i_\mu = \eta^\alpha F_{\mu\alpha}^i(A).
 \ee
 Here $\eta^\mu$ is the vector field (of mass dimension minus one) - generator of an infinitesimal diffeomorphism, and $F^i_{\mu\nu}(A)$ is the curvature of $A_\mu^i$. It can be checked that the above formula is a diffeomorphism corrected by a gauge transformation. Replacing the background curvature by its value (\ref{curv-0}) we get the following formula for an infinitesimal variation 
 \be
 \delta_\eta a^i_\mu = M_0^2 \eta^\alpha \Sigma^i_{\mu\alpha}.
 \ee
 This suggests that we consider vector fields $\xi^\mu =M_0^2 \eta^\mu$ of mass dimension one that are finite in the limit $M_0 \to 0$. Thus, let us consider the following variations
 \be\label{diffeo}
 \delta_\xi a^i_\mu = \xi^\alpha \Sigma^i_{\mu\alpha},
 \ee
 which will play the role of an infinitesimal diffeomorphism for the theory (\ref{lin-action}). 
 
 Another set of transformations that we have to consider are gauge symmetries. An infinitesimal gauge transformation is given by
 \be\label{gauge}
 \delta_\phi a^i_\mu = \partial_\mu \phi^i.
 \ee
 
 Let us now verify that the linearized action is invariant under (\ref{diffeo}) and (\ref{gauge}). For this we will need the following identity
 \be\label{ident}
\Sigma^{i\,\mu\nu} \Sigma^j_{\nu\rho} = -\delta^{ij} \eta^\mu{}_\rho + \epsilon^{ijk} \Sigma^{k\,\mu}{}_\rho\, ,
\ee 
which can be verified by a direct computation. Here $\eta^{\mu\nu}$ is the Minkowski metric. Let us first consider diffeomorphisms. Thus, consider the quantity
\be
\Sigma^{i\,\mu\nu} \partial_\mu \delta_\xi a^j_\nu =\Sigma^{i\,\mu\nu} \partial_\mu \xi^\alpha \Sigma^j_{\nu\alpha}.
\ee
Using (\ref{ident}) we see that $ij$-symmetric part of this quantity is proportional to $\delta^{ij}$. However this, when contracted with the projector in (\ref{lin-action}) gives zero. Thus, the invariance under infinitesimal changes of coordinates is established. The invariance under gauge transformations (\ref{gauge}) follows by noting that the quantity $\Sigma^{i\,\mu\nu}$ is anti-symmetric and therefore $\Sigma^{i\,\mu\nu} \partial_\mu \delta_\phi a^j_\nu=0$.

Since our gauge theory action (\ref{lin-action}) is both diffeomorphism and gauge invariant we can already make a suspected count of the number of propagating DOF. Indeed, the configurational variable of the theory should be the spatial projection of the connection. This has $3\times 3=9$ components. Subtracting 4 diffeomorphisms and 3 gauge DOF leaves us with 2 suspected propagating DOF. Let us confirm this count by the Hamiltonian analysis of the linearized theory. This will also help us to see the gravitons explicitly.

\subsection{Hamiltonian analysis}

In this subsection we give a more detailed demonstration of the spin two nature of our theory given in the introduction.

To obtain the action in the Hamiltonian form let us expand the quantity that appears as the main building block of the linearized action (\ref{lin-action}). We have
\be\label{ham-1}
\Sigma_i^{\mu\nu} \partial_\mu a^j_\nu = \im \partial_i a_0^j - \im \dot{a}_i^j - \epsilon_i^{kl} \partial_k a_l^j.
\ee
Here we have identified the spatial $a$ and internal $i$ indices using e.g. the component $\delta_a^i:=\Sigma_{0a}^i$ of the background two-form, and $\partial_i$ are the partial derivatives with respect to spatial coordinates. We raise and lower spatial indices freely using $\delta^{ij}$ metric. 

It is now easy to compute the conjugate momenta. Since the time derivatives that appear in the action are those of the spatial projection of the connection, it is clear that only these components can have non-zero momenta. However, since the projector is involved in (\ref{lin-action}), we see that only the symmetric tracefree part of $a_i^j$ has non-zero momenta. These are
\be\label{pi-conn}
\pi^{ij} = P^{ij|kl} \left( \dot{a}_{kl} - \partial_k a_{0\, l} - \im \epsilon_{kmn} \partial_m a_{n\, l} \right).
\ee
We note that the action (\ref{lin-action}) does not at all depend on the trace part of the spatial connection $a_i^j$. However, there is a dependence on the anti-symmetric (and of course symmetric) parts. Let us separate the trace, symmetric and anti-symmetric parts of $a_i^j$ and write
\be\label{conn-decomp}
a_{ij} = a_{ij}^s  + b\delta_{ij} + \epsilon_{ijk} c_k .
\ee
Here $a^s_{ij}$ is the symmetric and tracefree part, and $b, c_i$ parameterise the trace and anti-symmetric parts respectively. Let us now rewrite the expression for the momentum using this decomposition. We have
\be
\pi^{ij} = {\dot{a}}^{s\, ij} - \im \epsilon^{ikl} \partial_k a_{l}^{s\, j} + P^{ij|kl} \partial_k (\im c_l-a_{0\, l}). 
\ee
We note that the second term here is automatically symmetric and tracefree. On the other hand, it is clear that the Lagrangian density in (\ref{lin-action}) is 
\be
{\cal L}= \frac{(\pi^{ij})^2}{2}.
\ee
We see that the Lagrangian (density) is independent of $b$. This has a simple interpretation. Indeed, computing the infinitesimal diffeomorphism action on the temporal and spatial projections of the connection we find
\be
\delta a_0^i = \im \xi^i, \qquad \delta_\xi a^i_j = -\im \xi^0 \delta^i_j - \epsilon^i{}_{jk} \xi^k.
\ee
This in particular means that the trace part $b$ of the matrix $a_i^j$ is a pure gauge quantity that can be set to zero by a temporal diffeomorphism. We also see that the Lagrangian depends on the anti-symmetric part of spatial and temporal components of the connection only in the combination $\im c_i - a_{0\, i}$. Indeed, it is easy to check that precisely this combination is invariant under spatial diffeomorphisms, as the anti-symmetric component transforms as $\delta_\xi c_i = \xi_i$.  Let us denote the invariant combination by $\phi_i$. As we shall soon see, it will become a generator of infinitesimal gauge rotations in our theory. Thus, we finally rewrite the momentum as
\be
\pi^{ij} = {\dot{a}}^{s\, ij} - \im \epsilon^{ikl} \partial_k a_{l}^{s\, j} + P^{ij|kl} \partial_k \phi_l,
\ee
and compute the Hamiltonian density as ${\cal H}=\pi^{ij} \dot{a}_{ij}^s -{\cal L}$. We get
\be
{\cal H}=\frac{(\pi^{ij})^2}{2}+\im \pi^{ij} \epsilon_i{}^{kl}\partial_k a_{lj}+\phi_i \partial_j \pi^{ij},
\ee
where we have dropped the index $s$ from $a^s_{ij}$ for brevity. Thus, now all the dynamical fields appearing in the Hamiltonian are symmetric tracefree tensors. The quantity $\phi_i$ is the Lagrange multiplier, which serves as a generator of ${\rm SU}(2)$ rotations on the connection. Indeed, the Poisson bracket of the integrated last term with the connection gives
\be
\delta_\phi a_{ij} = \partial_{(i} \phi_{j)},
\ee
which is just the (symmetrised) gauge transformation. To see the structure of the arising Hamiltonian it is convenient to fix the gauge and require the connection to be transverse 
\be
\partial^i a_{ij}=0.
\ee
The momentum is required to be transverse by the condition obtained varying the action with respect to the Lagrange multipliers $\phi_i$. So, it is now clear that the reduced phase space of our linearized system is parameterised by two symmetric, tracefree and transverse matrices $a_{ij}$ and $\pi_{ij}$. This corresponds to two propagating DOF.

Let us now see what the dynamics becomes. To unravel the structure of the arising expression for the (reduced) Hamiltonian let us further rewrite it as
\be\label{ham}
{\cal H}=\frac{1}{2}(\pi^{ij}+ \im \epsilon^{ikl}\partial_k a_l^j)^2 + \frac{1}{2}(\partial_k a_{ij})^2.
\ee
Up to this point no reality conditions for the fields were specified. We can now deduce the linearized theory reality conditions from the Hamiltonian (\ref{ham}). Indeed, declaring the symmetric tracefree transverse connection field $a_{ij}$ to be real, and defining a new real momentum field
\be
p^{ij}:= \pi^{ij}+ \im \epsilon^{ikl}\partial_k a_l^j, \qquad p^{ij}\in \R
\ee
we can rewrite the linearized Hamiltonian in an explicitly positive definite form
\be
{\cal H} = \frac{1}{2}(p^{ij})^2 + \frac{1}{2}(\partial_k a_{ij})^2.
\ee
The field equations that follow are now the usual 
\be
\Box \, a_{ij} = 0,
\ee
which is just the wave equation for the two components of the connection field $a_{ij}$. This is how gravitons are described by our gauge theory approach. We note that one can recognise in the analysis of this section the linearized version of the new Hamiltonian formulation of gravity \cite{Ashtekar:1991hf}. In particular, the arising reality conditions for the phase space fields are  the same as in this formulation. Thus, even though our starting point of a gauge theory is a bit unconventional, the linearized theory mimics constructions familiar from other formulations. 

What is different about our linearized theory (\ref{lin-action}) from the more familiar treatment in \cite{Ashtekar:1991hf} is that no diffeomorphism constraints are left in the final result. Instead, our linearized action is simply independent of certain components of the connection field, so the theory is formulated on a smaller configuration space to start with. In other words, in our pure connection approach to gravity the Hamiltonian and diffeomorphism constraints of GR that usually require so much attention are solved once and for all by projecting out certain components of the connection field. This fact about our formulation must be very important for practical applications. And indeed, we shall see below that e.g. the issue of the gauge-fixing is considerably easier here than in the case of metric based GR.

 \section{Propagator}
 \label{sec:prop}
 
 In this section we invert the quadratic form that we have obtained by expanding the theory around the Minkowski spacetime background. In doing this we must decide on the gauge fixing. 
 
 \subsection{Gauge fixing}
 
 We have seen that the action (\ref{lin-action}) is invariant under both gauge and diffeomorphism transformations, but we have also seen above that this invariance is manifested very differently in the two cases. Thus, in the case of the gauge invariance the situation is completely standard in that some of the field components have zero momenta and are thus Lagrange multipliers --- generators of gauge symmetries. In the case of diffeomorphisms the situation is very different --- we have seen that the action is simply independent of some components of the field, exactly those components that can be freely changed by performing a diffeomorphism. Thus, while there is very little choice for dealing with the gauge rotations --- we have to treat them in the usual way by fixing the gauge and thus making the unphysical components of the gauge field propagate --- we will need a different procedure for dealing with those components of the connection that gets affected by diffeomorphisms. 
 
 A useful analogy here is as follows. Let us consider a theory of two scalar fields $\phi, \psi$ with the Lagrangian
 \be
 {\cal L} = - \frac{1}{2}(\partial_\mu(\phi-\psi))^2.
 \ee
 It is clear that the Lagrangian is invariant under a simultaneous shift of both of the fields by some function. The way this is realized is that the Lagrangian is simply independent of a certain combination of the fields, namely of $\phi+\psi$, being only a function of the combination $\phi-\psi$. A natural quantization strategy in this case is to introduce a new field $\phi-\psi$ and rewrite the Lagrangian in terms of the new field only. Then only this combination of the fields is a propagating field, while the other combination $\phi+\psi$ is a fiction.
 
 In the case of the simple Lagrangian above it is very easy to see what the propagating field is. In our case (\ref{lin-action}) this is much harder. In particular, we will not be able to rewrite the full Lagrangian in a way that has only diffeomorphism invariant combinations of the connection components appearing (see, however, below for an expression for the linearized Lagrangian that depends solely on the "physical" diffeomorphism invariant components of the connection). However, an appropriate strategy is as follows. We can consider the quadratic form (\ref{lin-action}) as a form on the space of diffeomorphism invariant classes of connections $a_\mu^i$, i.e. connections related via
 \be\label{sim-diff}
 a_\mu^i \sim a_\mu^i + \xi^\nu \Sigma^i_{\mu\nu}.
 \ee
 The quadratic form in (\ref{lin-action}) is degenerate on this space because there is still the usual gauge invariance to be taken care of. However, this gauge invariance can be dealt with in the usual way, by fixing the gauge. As we shall see below, it will be possible to find a gauge-fixing condition that is invariant under (\ref{sim-diff}). After doing this we obtain a non-degenerate quadratic form on the space of diffeomorphism classes (\ref{sim-diff}). It can be inverted, to obtain a propagator on the space of diffeomorphism classes of connections. As is standard for gauge-fixing, this procedure will make the temporal and longitudinal components of the connection propagate (and will add ghosts that will offset the effect of making this components propagating). At the same time, the components of the connection that are identified in (\ref{sim-diff}) will not be propagating, as the propagator will involve a projector on the space of diffeomorphism equivalence classes. This way of dealing with the gauge symmetries of our theory is very different from the case of the metric based GR, but is quite natural given that the diffeomorphisms are realized in our theory quite differently. 
  
 Having explained the logic of our procedure it remains to find a gauge-fixing condition that is diffeomorphism invariant. After some trial and error we found the following gauge-fixing condition to be useful:
 \be\label{gf-cond}
 \partial^\mu \Pi^{\mu i| \nu j} a_{\nu j} = \frac{2}{3} \partial^\mu \left(a_\mu^i +  \frac{1}{2} \epsilon^{ijk} \Sigma^k_\mu{}^\nu a_\nu^j\right)=0,
 \ee
where
 \be\label{proj-P}
\Pi^{\mu i| \nu j} := \eta^{\mu\nu}\delta^{ij} + \frac{1}{3} \Sigma^{i\,\mu\rho} \Sigma_\rho^j{}^\nu = \frac{2}{3} \left(\eta^{\mu\nu} \delta^{ij} +\frac{1}{2} \epsilon^{ijk}\Sigma^{k\,\mu\nu} \right)
\ee
is a projector operator whose meaning is to be clarified below. The projector property
\be
\Pi^{\mu i| \nu j} \Pi_{\nu j}{}^{\rho k} = \Pi^{\mu i| \rho k},
\ee
can be checked by an elementary computation. It is easy to see that our gauge-fixing condition is diffeomorphism invariant. Indeed, consider
 \be\label{gf-1}
  \xi^\nu \Sigma^i_{\mu\nu} + \frac{1}{2} \epsilon^{ijk} \Sigma^k_\mu{}^\nu \xi^\rho \Sigma^j_{\nu\rho}.
  \ee
  Using the algebra (\ref{ident}) of $\Sigma^i_{\mu\nu}$ matrices we see that the last term here equals
  \be
  \frac{1}{2} \epsilon^{ijk} \epsilon^{kjl} \xi^\rho \Sigma^l_{\mu\rho} = - \xi^\nu \Sigma^i_{\mu\nu} .
  \ee
  Thus, the quantity in (\ref{gf-1}) is zero, and the gauge-fixing condition (\ref{gf-cond}) is diffeomorphism-invariant. It is also clear that as far as the gauge transformations are concerned the last term in (\ref{gf-cond}) is inessential, for it is zero for any $a_\mu^i$ that is a pure gauge $a_\mu^i=\partial_\mu \phi^i$. Thus, (\ref{gf-cond}) is the usual gauge theory gauge-fixing condition, corrected by a term that is inessential as far as the behaviour under the gauge transformations is concerned. 
  
Let us now confirm that the projector $\Pi^{\mu i| \nu j}$ is just that on diffeomorphism equivalence classes of connections, and so it is natural to apply it before the usual gauge-fixing condition is imposed (to make this condition diffeomorphism invariant). We compute the action of the projector on the connection $a_\nu^j$ decomposed as in the previous subsection
\be
a_\nu^j = a_0^j (dt)_\nu + (a_{ij}^s + b \delta_{ij} + \epsilon_{ijk} c_k)(dx^i)_\nu.
\ee
The result is
\be
\Pi^{\mu i| \nu j} a_\nu^j = \frac{2}{3}\left( \delta^{ij} \left(\frac{\partial}{\partial t}\right)^\mu + \frac{i}{2}\epsilon^{ijk} \left( \frac{\partial}{\partial x^k}\right)^\mu \right) (a_0^j-\im c^j) + a_{ij}^s \left( \frac{\partial}{\partial x^j}\right)^\mu.
\ee
We note the the quantity $b$ got projected out, and the projected connection only depends on the temporal and the anti-symmetric spatial components of the connection in the combination $a_0^i-\im c^i$, as expected from the previous section. Thus, the projector $\Pi^{\mu i| \nu j}$ is indeed just that on the diffeomorphism invariant subspace, and selects the components $a_0^i-\im c^i$, which play the role of the generators of the Gauss constraints, as well as $a_{ij}^s$, which are the two propagating DOF plus three longitudinal modes of the connection. As usual for a gauge theory we shall make the components generators of the Gauss constraints as well as the longitudinal components of the connection propagating by adding a gauge fixing term, and then offset their effects by adding ghosts. 

The projector $\Pi^{\mu i|\nu j}$ can be somewhat demystified by explaining what is its spinorial analog. Readers not familiar with the Penrose's spinor language \cite{Penrose:1985jw} for gravity can skip this paragraph. Using spinors one can express the connection $a_\mu^i$ as a certain rank 4 spinor. Indeed, the spacetime index gets replaced by a pair $AA'$ of an unprimed and primed spinor indices. The ${\rm SU}(2)$ index $i$ gets replaced by a pair $AB$ of two unprimed indices, which is moreover $AB$ symmetric. Thus we get $a^{AB}{}_{CC'}$ as our linearized theory dynamical field. It is now not hard to show that the projector $\Pi^{\mu i|\nu j}$ is simply that on the component of this spinor that is completely symmetric in all its 3 unprimed indices. Thus, schematically, $(\Pi a)_{AB\, CC'} = a_{(ABC)C'}$, where the brackets denote symmetrization. The projected out part is $a^{AB}{}_{BA'}$, and is thus a mixed rank 2 spinor and carries precisely 4 components, as is appropriate for something that can be projected out by a diffeomorphism. As we shall note below, the linearized action (\ref{lin-action}) can be written very simply in terms of the field $a_{(ABC)C'}$.

 We now add the gauge-fixing condition squared with some parameter to the Lagrangian. Thus, we consider the following gauge-fixed Lagrangian on the space of diffeomorphism equivalence classes of connections
  \be\label{L-gf}
  {\cal L}_{\rm gf} = -\frac{1}{2} P^{ij|kl} (\Sigma^{i\,\mu\nu} \partial_\mu a_\nu^j) (\Sigma^{k\,\rho\sigma} \partial_\rho a_\sigma^l) 
  -\frac{\alpha}{2} \left(  \partial^\mu a_\mu^i -\frac{1}{2} \epsilon^{ijk} \Sigma^{j\,\mu\nu} \partial_\mu a_\nu^k \right)^2,
  \ee
  where we have changed the order of indices $jk$ in the gauge-fixing term for convenience, and absorbed the $(2/3)^2$ factor into the gauge-fixing parameter $\alpha$. As in the case of Yang-Mills theory, the idea is now to select the gauge-fixing parameter $\alpha$ so that the gauge-fixed action is as simple as possible. 

\subsection{The algebra of gauge-fixing}

In this subsection we will simplify the expression for the gauge-fixed Lagrangian and find a useful value for the gauge-fixing parameter $\alpha$. To this end, let us first write the Lagrangian in the momentum space. Omitting the argument $\pm k$ from the Fourier components $a_\mu^i(k)$ of $a_\mu^i$ for brevity we have the following expression
\be
 {\cal L}_{\rm gf} = -\frac{1}{2} P^{ij|kl} (\Sigma^{i\,\mu\nu} k_\mu a_\nu^j) (\Sigma^{k\,\rho\sigma} k_\rho a_\sigma^l) 
  -\frac{\alpha}{2} \left(  k^\mu a_\mu^i -\frac{1}{2} \epsilon^{ijk} \Sigma^{j\,\mu\nu} k_\mu a_\nu^k \right)^2.
  \ee
  Let us expand the last term. Introducing a compact notation $(ka^i):=k^\mu a_\mu^i$ and expanding the product of two $\epsilon$'s we have
  \be\label{gf-2}
  \left(  (ka^i) -\frac{1}{2} \epsilon^{ijk} \Sigma^{j\,\mu\nu} k_\mu a_\nu^k \right)^2 =(ka^i)^2 - (ka^i)\epsilon^{ijk} \Sigma^{j\,\mu\nu} k_\mu a_\nu^k \\ \nonumber
  +\frac{1}{4} \Sigma^{i\mu\nu} \Sigma^{i\,\rho\sigma} k_\mu k_\rho a_\nu^j a_\sigma^j - 
  \frac{1}{4} (\Sigma^{i\mu\nu} k_\mu a_\nu^j) (\Sigma^{j\,\rho\sigma}  k_\rho  a_\sigma^i) .
  \ee
  
  Let us now expand the first term of the Lagrangian. We have
  \be\label{gf-3}
  P^{ij|kl} (\Sigma^{i\,\mu\nu} k_\mu a_\nu^j) (\Sigma^{k\,\rho\sigma} k_\rho a_\sigma^l) =
  \frac{1}{2} \Sigma^{i\mu\nu} \Sigma^{i\,\rho\sigma} k_\mu k_\rho a_\nu^j a_\sigma^j
 \\ \nonumber +\frac{1}{2} (\Sigma^{i\mu\nu} k_\mu a_\nu^j) (\Sigma^{j\,\rho\sigma}  k_\rho  a_\sigma^i)
  -\frac{1}{3} (\Sigma^{i\,\mu\nu} k_\mu a_\nu^i) (\Sigma^{j\,\rho\sigma}  k_\rho  a_\sigma^j).
  \ee
  We can now use the following two identities
  \be\label{ident-proj}
  \Sigma^{i\mu\nu} \Sigma^{i\,\rho\sigma} = \eta^{\mu\rho} \eta^{\nu\sigma}-  \eta^{\nu\rho} \eta^{\mu\sigma}-\im \epsilon^{\mu\nu\rho\sigma}
  \ee
  and
  \be\label{ident-as}
  \Sigma^{i\mu\nu} \Sigma^{j\,\rho\sigma} - \Sigma^{j\mu\nu} \Sigma^{i\,\rho\sigma} =
  \epsilon^{ijk} \left( \Sigma^{k\, \mu\sigma} \eta^{\nu\rho} - \Sigma^{k\, \nu\sigma} \eta^{\mu\rho} - \Sigma^{k\, \mu\rho} \eta^{\nu\sigma}+\Sigma^{k\, \nu\rho} \eta^{\mu\sigma}\right).
  \ee
  We can now use the identity (\ref{ident-proj}) to rewrite the first term in (\ref{gf-3}), and the identity (\ref{ident-as}) to rewrite the last term as a multiple of the second plus some extra terms. We get
  \be
  \frac{1}{2}(k^2 (a_\mu^i)^2 - (ka^i)^2) + \frac{1}{6} (\Sigma^{i\mu\nu} k_\mu a_\nu^j) (\Sigma^{j\,\rho\sigma}  k_\rho  a_\sigma^i)
  +\frac{1}{3} \left( k^2 \epsilon^{ijk} \Sigma^{i\,\mu\nu} a_\mu^j a_\nu^k+ 2(ka^i) \epsilon^{ijk} \Sigma^{j\,\mu\nu} k_\mu a_\nu^k\right).
  \ee
  We now note that if we make a choice
  \be
  \alpha = \frac{2}{3}
  \ee
  then the terms $(\Sigma^{i\mu\nu} k_\mu a_\nu^j) (\Sigma^{j\,\rho\sigma}  k_\rho  a_\sigma^i)$, as well as $(ka^i)^2$  and $(ka^i) \epsilon^{ijk} \Sigma^{j\,\mu\nu} k_\mu a_\nu^k$ cancel out and we get the following simple gauge-fixed action
  \be\label{gf*}
  {\cal L}_{\rm gf} = -\frac{k^2}{3}\left( (a_\mu^i)^2 +\frac{1}{2} \epsilon^{ijk}\Sigma^{k\,\mu\nu} a_\mu^i a_\nu^j \right)=-\frac{k^2}{2}\Pi^{\mu i| \nu j} a_{\mu i} a_{\nu j},
  \ee
  where $\Pi^{\mu i| \nu j}$ is the projector (\ref{proj-P}). Because the projector on diffeomorphism equivalence classes appears here explicitly, it is obvious that this action is still invariant under the diffeomorphisms (\ref{sim-diff}), and so is now a non-degenerate quadratic form on the space of diffeomorphism equivalence classes.
  
  We note that the above analysis implies that our original linearized Lagrangian (\ref{lin-action}) can be rewritten (in the momentum space) in terms of the "projected" connection $\Pi a$ schematically as follows:
  \be\label{Lagr-YM}
  {\cal L}= -\frac{1}{2}\left( k^2 (\Pi a)^2 - \frac{3}{2} (k \Pi a)^2 \right).
  \ee
  Thus, our linearized Lagrangian is {\it different} from that for Yang-Mills theory for the projected connection $\Pi a$. Indeed, in the case of Yang-Mills the numerical coefficient in front of the second term in the brackets in (\ref{Lagr-YM}) would be unity. In the case of Yang-Mills theory the value of the coefficient in front of $(ka)^2$ is fixed by the requirement of gauge invariance. The same is true in our case, and the different numeric value has to do with the fact that the projected connection $\Pi a$ transforms under the gauge transformations in a more complicated way than $\delta a_\mu^i =\partial_\mu \phi^i$. Indeed, we have:
  \be
  \delta \Pi^{\mu i|\nu j} a^j_\nu = \Pi^{\mu i|\nu j} \delta a_\nu^j = \frac{2}{3} \partial_\mu\phi^i - \frac{1}{3}\epsilon^{ijk} \Sigma_\mu^{\,\,\nu j} \partial_\nu \phi^k.
  \ee
 It is this more involved transformation law for the projected connection that is responsible for the different from the Yang-Mills case numerical factor in front of the second term in (\ref{Lagr-YM}). 
 
 We can now also note that our linearized Lagrangian in (\ref{lin-action}) admits a very simple description in terms of spinors. Thus, as we have already mentioned, in the spinor notation our connection $a_\mu^i$ gets described by a rank 4 spinor $a_{ABCC'}$. The diffeomorphism classes are described by the component which is symmetric in its 3 unprimed indices, or, in other words, by the $(3/2,1/2)$ irreducible representation of the Lorentz group, where the first number denotes the representation in the space of unprimed spinors and the second one in the space of primed ones. The Lagrangian in (\ref{lin-action}) is then a multiple of
 \be\label{L-spinor}
 {\cal L}\sim (\partial^{(A}{}_{A'} a^{BCD)A'})^2,
 \ee
 where the precise numerical coefficient is convention dependent and will be spelled out elsewhere. Here $\partial_{AA'}$ is the 2-component spinor Dirac operator. In words, the Dirac operator is used to convert the representation $(3/2,1/2)$ described by the connection to the spin 2 representation $(2,0)$, and this is then squared to form the Lagrangian. The Lagrangian clearly only depends on the part $a_{(ABC)C'}$ of the connection, which makes it obvious that at least in the linearized theory the diffeomoprhisms are realized simply so that the action is independent of some of the connection components. The form (\ref{L-spinor}) of the Lagrangian also explains the structure of the propagator that is obtained below. 

\subsection{Propagator}

We now invert the quadratic form in (\ref{gf*}). Thus, we add a current term to the action
\be
S_{\rm gf} = \int \frac{d^4 k}{(2\pi)^4}\left[ - \frac{k^2}{2} \Pi^{\mu i| \nu j} a_{\mu i}(-k) a_{\nu j}(k) + J^{\mu\,i}(-k)a_\mu^i(k)\right],
\ee
and then integrate the field $a_\mu^i$ out. This can be easily done in the space of diffeomorphism equivalence classes, and we immediately see that the action with the original connection field integrated out is given by
\be
S[J] = \int \frac{d^4 k}{(2\pi)^4} \frac{1}{2k^2} \Pi^{\mu i| \nu j} J_\mu^i(-k) J_\nu^j(k).
\ee
In other words, the propagator of our theory is given by
\be
\langle a^{\mu i}(-k) a^{\nu j}(k) \rangle = (1/\im)^2 \frac{\delta}{\delta J_\mu^i(-k)} \frac{\delta}{\delta J_\nu^j(k)} 
e^{\im S[J]} \Big|_{J=0} = (1/\im) \frac{1}{k^2} \Pi^{\mu i| \nu j},
\ee
which is just the usual $1/k^2$ term times the projector onto the space of diffeomorphism equivalence classes of connections, times the (convention dependent) $1/\im$ factor.

This finishes our discussion of the free theory of gravitons on the Minkowski spacetime background (or gravitons with energy $E\gg M_0$ much greater than the energy scale of our constant curvature background). We refrain from considering ghosts that are irrelevant for our purely classical purposes in this paper. Instead, let us now consider the lowest order interactions.

\section{Interactions}
\label{sec:inter}

In this section we consider graviton self-interactions and discuss puzzles related to the fact that the action blows up in the $\Lambda\to 0$ limit.

\subsection{Third variation of the action}

The third variation of the action is easily computed from (\ref{sec-var}). We get
\be\label{3-var}
\delta^3 S[A] = (1/\im) \int_M d^4x\, \left( \frac{\partial^3 f}{\partial \tilde{X}^{ij}\partial \tilde{X}^{kl}\partial \tilde{X}^{pq}} \delta \tilde{X}^{ij} \delta \tilde{X}^{kl}  \delta \tilde{X}^{pq}
+3 \frac{\partial^2 f}{\partial \tilde{X}^{ij}\partial \tilde{X}^{kl}} \delta^2 \tilde{X}^{ij} \delta X^{kl} 
+ \frac{\partial f}{\partial \tilde{X}^{ij}} \delta^3 X^{ij}  \right).
\ee

We have already computed the first and second variations of the matrix $\tilde{X}^{ij}$ in (\ref{X-1}), (\ref{X-2}). The third variation is given by
\be\label{X-3}
\delta^3 \tilde{X}^{ij} = \frac{3}{2}\tilde{\epsilon}^{\mu\nu\rho\sigma} D_{A\, \mu} \delta A^{(i}_\nu \epsilon^{j)kl} \delta A_\rho^k \delta A_\sigma^l.
\ee
We also note that the fourth variation, of relevance for higher-order interaction vertices, is zero, which follows by expanding the product of two $\epsilon$'s and noting that there is always a $\delta^{ij}$-contraction of two variations of the connection. On the other hand, spacetime indices of all 4 variations of the connection are contracted with $\tilde{\epsilon}^{\mu\nu\rho\sigma}$, and so the result is zero.

\subsection{Cubic interaction}

We have already computed the first and second derivatives of the defining function at the identity matrix in (\ref{fp}), (\ref{fpp}). Let us now compute the third derivative. Here we only consider a simpler case when the defining function depends on the invariant ${\rm Tr}(X^2)/({\rm Tr})^2$. The general case will be described elsewhere. We get:
\be
 \frac{\partial^3 f}{\partial X^{ij} \partial X^{kl} \partial X^{pq}} \Big|_{X_0} = -\frac{2g_0}{3(-2\im M_0^4)^2} \left( \delta^{ij} P^{kl|pq} + \delta^{kl} P^{ij|pq} + \delta^{pq} P^{ij|kl}\right),
 \ee
where $g_0$ is the dimensionless constant given by (\ref{g}), and $P^{ij|kl}$ is the projector on the symmetric traceless part that we already encountered above. 

We now compute the cubic interaction term. Let us first discuss the simpler case when all 3 gravitons are high energy $E\gg M_0$. In this case certain terms are dominant, and we are first going to describe these terms. We evaluate (\ref{3-var}) at the constant curvature background connection (\ref{A0}). The last term in (\ref{3-var}) is then seen to be a total derivative. We can also note that of the two terms coming from $\delta^2 X^{ij}$ one term is proportional to $(D\delta A)^2$, while the other is of the order $M_0^2 (\delta A)^2$. Let us first neglect the term $M_0^2 (\delta A)^2$ as compared to $(D\delta A)^2$. Then, after some rewriting we get
\be
\delta^3 S \Big|_{A_0} = \frac{g_0}{2 M_0^2} \int d^4x \sqrt{-g}\, P^{ij|kl} (\Sigma^{i\,\mu\nu} D_{A_0\,\mu} \delta A_\nu^j) \Big[ (\Sigma^{k\,\rho\sigma} D_{A_0\,\rho} \delta A_\sigma^l)  (\Sigma^{m\, \alpha\beta}D_{A_0\,\alpha} \delta A_\beta^m) \\ \nonumber-3\im \, \epsilon^{\alpha\beta\gamma\delta} D_{A_0\,\alpha} \delta A_\beta^k \, D_{A_0\,\gamma} \delta A_\delta^l \Big]  .
\ee
Now passing to the high-energy limit $E\gg M_0$ we replace the covariant derivatives by the usual coordinate ones, and then rewrite the interaction term in terms of the connection field $a_\mu^i= \sqrt{g_0} (\delta A_\mu^i)$, for which the kinetic term (\ref{lin-action}) is canonically normalised. We also need to divide the third variation by $3!$ to get the correct (leading contribution to) the cubic interaction term. We get:
\be
S^{(3)} = \frac{1}{12 \sqrt{g_0} M_0^2} \int d^4x \, P^{ij|kl} (\Sigma^{i\,\mu\nu} \partial_\mu a_\nu^j) \Big[ (\Sigma^{k\,\rho\sigma} \partial_\rho a_\sigma^l)  (\Sigma^{m\, \alpha\beta}\partial_\alpha a_\beta^m) -3\im \, \epsilon^{\alpha\beta\gamma\delta} \partial_\alpha a_\beta^k \partial_\gamma a_\delta^l \Big].
\ee
To summarize, schematically, the obtained leading order contribution to the cubic interaction is of the form
\be\label{inter}
{\cal L}^{(3)} \sim \frac{1}{\sqrt{g_0} M_0^2} (\partial a)^3.
\ee
We learn that our theory of gravity has a negative mass dimension coupling constant, and so is non-renormalisable in the usual sense of the word, as could be expected. We also see that in our approach the self-coupling of our gravitons described by the connection perturbation $a$ cannot be identified with the Newton's constant. Indeed, for the defining function (\ref{def-GR}) that corresponds to the cosmological constant GR we have $g_0\sim M_p^2/M_0^2$. Thus, we see that the combination that appears in the denominator of (\ref{inter}), at least for the defining function that corresponds to GR, is given by 
\be\label{Mp-M0}
M_*^2 := \sqrt{g_0} M_0^2 \sim M_p M_0.
\ee

\subsection{Discussion}

Some remarks on the result (\ref{inter}) are in order. First, the obtained form of the cubic graviton self-interaction is different from that in GR. Indeed, the GR Lagrangian expanded (around the Minkowski metric) starts with the cubic interaction term $\kappa h (\partial h)^2$, where $\kappa\sim \sqrt{G}\sim 1/M_p$ and $h$ is the metric perturbation. The GR cubic term is quadratic in the derivative operator, while (\ref{inter}) is cubic. This explains why the mass dimension of the coupling constant in (\ref{inter}) is minus two while in the cubic interaction term of GR it is minus one. Thus, both are non-renormalisable by power counting, but the form of the interaction is different. 

We also see that the coupling constant measuring the strength of self-interactions of gravitons in our approach is different from that in GR. This is not too surprising since in the usual metric-based approach the Newton's constant $G$ sets the strength of interaction of gravitons with the stress-energy of matter (or other gravitons). This is why it is a factor of $\sqrt{G}$ that serves as the theory's coupling constant. But the notion of the stress-energy tensor of gravitons is metric based. Indeed, the stress-energy arises as the variational derivative of the action with respect to the metric. In our approach gravitons are described in terms of a different field (connection $a$), and so the variation of the action with respect to $a$ no longer has the meaning of the stress tensor. This is why the strength of self-interaction of the connection field $a$ no longer needs to be directly identified with the Newton's constant. 

This raises the question of how the Newton's constant can be identified in our theory. One way to do this could be to evaluate the 4-graviton scattering amplitude, which in the usual metric based approach is proportional to $G$. We leave this calculation to future work.

Another remark on (\ref{inter}) is that $M_*$ given by (\ref{Mp-M0}) is the scale at which our perturbation theory appears to become strongly coupled. Thus, it appears that, unlike in the case of GR where the cutoff scale is $M_p$, the cutoff for the gravitational perturbation theory in the "pure connection" formulation is $M_*$. For the currently accepted value of the cosmological constant this is $M_*\sim 10^{-2} eV$. Thus, it appears that our perturbation theory cannot be trusted for energies larger than $10^{-2} eV$. While this fact would not be a problem for the envisaged renormalization group calculations that are non-perturbative in nature, this apparent strong coupling arising in our theory at such a low energy scale should certainly be given an interpretation. This is in particular worrying given the fact that for a particular defining function our theory is claimed to be the usual GR, with its very different strong coupling scale. There is clearly a puzzle here. 

While we have not yet worked out a resolution of this puzzle in all details, we believe what happens is as follows. The first remark is that (\ref{inter}) is not the full cubic vertex, but only its part that blows up in the $M_0\to 0$ limit. It is not hard to see that we have neglected another part (which goes to zero in the $M_0\to 0$ limit) and that the full cubic interaction term is schematially
\be\label{inter-full}
{\cal L}^{(3)}_{\rm full} \sim \frac{1}{\sqrt{g_0} M_0^2} (\partial a)^3 + \frac{M_0}{\sqrt{g_0}} (\partial a) a^2.
\ee
Thus, the cubic vertex consists of two parts. One blows up in the limit $M_0\to 0$, which is not very surprising given that the action itself blows up in this limit (at least in the case of GR when we can identify $M_0^2$ with $\Lambda$). The other goes to zero in the same limit. The fact that there is a blowing up part seems to indicate that it is not possible to take the Minkowski spacetime limit, which would be very worrying given that we certainly would like to be able to scatter gravitons in Minkowski spacetime to be able to compare predictions of our theory to those of the usual metric based formulation. However, it can be shown that the full interaction vertex (\ref{inter-full}) actually vanishes when all 3 external legs are put on shell. This is the same result as in GR, so in spite of some off-shell blowing up terms the on-shell result is completely the same as in GR. 

Thus, to understand what happens one must consider higher order interactions. One finds that the quartic interaction is schematically 
\be
{\cal L}^{(4)} \sim \frac{1}{g_1 M_0^4} (\partial a)^4 +  \frac{1}{g_0 M_0^2} (\partial a)^2 a^2 + \frac{1}{g_0} a^4,
\ee
where $g_1$ is a new coupling constant, related to higher derivatives of the defining function computed at the identity matrix. We see that there is again a blowing up leading order term. The last term vanishes in the limit $\Lambda\to 0$, when $g_0\to \infty$. However, we see that (in the case of GR) the second term is exactly the usual $(1/M_p^2) (\partial a)^2 a^2$ second-derivative graviton interaction. Thus, we see that when the 4-graviton scattering amplitude is computed there is a blowing up contribution from the diagrams involving two cubic vertices, as well as another blowing up contribution from the quartic vertex. There are also finite contributions both from the quartic vertex as well as from the diagrams involving two cubic vertices. We believe that, when evaluated on the physical states, the blowing up contributions should cancel, while the finite pieces assemble into the usual GR result. We will not attempt such a calculation here as it requires technology (spinor helicity) that is beyond the scope of this paper. But the fact that the terms finite in the $\Lambda\to 0$ limit are precisely of the familiar from GR two-derivative form support the picture sketched.

To summarize, the structure of interactions in our gauge-theoretic description of gravity is yet to be unravelled. It is, however, clear that the theory is as non-renormalizable as the usual GR in the metric based approach. What is different about our formulation is that the limit of the cosmological constant going to zero is a non-trivial one to take, for the action of the theory blows up in this limit. This is manifested in the fact that the interaction vertices contain blowing up pieces. Naively, this suggests strong coupling at a very low energy scale. However, we believe that the issue is much more subtle and that when computed for the physical graviton states the scattering amplitudes are perfectly finite in the $\Lambda\to 0$ limit and for the case of the defining function corresponding to GR reproduce the known results. A verification of this is left to future work. 

\section{Conclusions}

In this paper we have proposed a new approach to the gravitational perturbation theory. While our main motivation was the quantum theory (renormalization), in the present paper we remained in the classical domain. We have recalled how a diffeomorphims invariant gauge theory can be formulated using a homogeneous degree one defining function, and how such a theory for the gauge group ${\rm SU}(2)$ is a gravity theory describing two propagating degree of freedom. In particular, general relativity itself can be put in this framework, see the action (\ref{sec-GR-action}). Our main interest here was in the perturbation theory. Hence, we expanded our general diffeomorphims invariant gauge theory Lagrangian around a constant curvature connection (\ref{A0}). The original theory does not have any dimensionful parameters, and we have seen that it is the choice of the background that brings in a dimensionful quantity into the game, in our case the radius of curvature of the background. We then took a limit of the radius of curvature becoming very large (or working at energies such that the curvature of the background can be neglected). This way we obtained a theory on the Minkowski spacetime background.

The linearized action (\ref{lin-action}) we obtained is quite simple, and can be seen to be a natural construct involving the linearized connection, as well as the basic (anti-) self-dual two-forms $\Sigma_{\mu\nu}^i$. Indeed, as is sometimes done in the literature, one can introduce the derivative operators $\partial^{\mu\,i} := \Sigma^{\mu\nu\, i} \partial_\nu$. The basic building block of our linearized action is then $\partial^{\mu\, i} a_\mu^j$, where this quantity is symmetrised and then its tracefree part is squared to form the action. Note that the projector $P^{ij|kl}$ on the symmetric tracefree part is just that on the spin two part of the tensor product of two spin one representations, and this is another manifestation of how the spin two appears in the game. Indeed, one could rewrite our linearized gauge theory action using the spinor notation as a multiple of $(\Sigma^{\mu\nu\,(AB}\partial_\mu a^{CD)}_\nu)^2$, where the brackets denote the symmetrisation. A completely symmetrised rank 4 spinor is the standard realisation of the spin two representation. Another, particularly clear way to rewrite our linearized Lagrangian is completely in terms of spinors, when all spacetime indices are eliminated in favour of the spinor ones. The Lagrangian then takes the extremely simple form (\ref{L-spinor}). This should be compared to a much more involved linearized Lagrangian for gravitons in the usual metric-based approach. This considerable simplification of the linearized Lagrangian is in itself a significant plus of our approach. 

Another very important feature of our approach is that diffeomorphism invariance can be dealt with in a very simple way. Recall that it is this gauge symmetry that is causing so much difficulty in any approach to gravity, perturbative or non-perturbative. In contrast, in our formulation diffeomorphisms can be dealt with once and for all, by simply projecting out certain components of the connection. We believe that this feature of our gauge-theoretic description of gravity is very important, to be fully appreciated with more work on this approach. In a certain cense, what replaced the usual diffeomorphisms in our approach are the ${\rm SU}(2)$ gauge rotations. We have seen that these must be gauge-fixed in the usual fashion. It is however much easier to deal with gauge rotations than with diffeomorphisms, something that can be appreciated from our derivation of the propagator of our theory. This propagator can be literally read off from the Lagrangian in its form (\ref{L-spinor}). There is no such a simple derivation of the propagator for the metric-based gravitons. 

We have also looked at the (cubic and quartic) graviton self-interactions as described in our gauge-theoretic framework. It was observed that the perturbation theory appears to become strongly coupled at a very low energy scale $M_*=\sqrt{M_p M_\Lambda}$, and so appears to behave quite differently from the perturbation theory based on the Einstein-Hilbert Lagrangian. However, there are indications that this strong coupling may be only apparent, and that the physical scattering amplitudes are the same as in the metric based GR. A verification that this is indeed the case is left to future work. 

Apart from quantum aspects, which we purposefully decided to avoid here, we did not comment much on the subtle issue of the reality conditions for our theory. Indeed, these were discussed at the linearized level, where their treatment is no different from that in the Ashtekar formulation, see \cite{Ashtekar:1991hf}. It is clear, however, that the full interacting action will require a much more sophisticated choice of the reality conditions. For the quantum calculations to be carried out with this formalism this is not much of an issue, because all loops are computed via the trick of the analytic continuation, and under this all factors of $\sqrt{-1}$ in our formulas disappear and fields become real. However, these issues do matter for the questions of the unitarity of the arising quantum theory. We expect that these subtle issues will take some time to be settled, and refrain from trying to address them in this work. 

To conclude, we hope to have convinced the reader that the present gauge-theoretic approach to gravity brings with itself many rather exciting opportunities that are simply unavailable, or impractical in the usual metric setting. It now seems within reach that, with the new tools developed here, the renormalization group flow for an infinite parametric class of gravity theories can be computed. Once this is achieved, ideas about the ultra-violet behaviour of gravity, e.g. the asymptotic safety conjecture \cite{Weinberg:2009bg}, can be explicitly tested. 

\section*{Acknowledgements} The author is grateful to  J. Fine and D. Panov for making a draft of the paper \cite{Joel} available to him prior to publication. Some of its constructions made the author take the "pure connection" formulation of diffeomorphism invariant gauge theories more seriously, which resulted in the present paper. 

\section*{Appendix A: Defining function for GR with the cosmological constant}

In this appendix we derive an expression for the defining function corresponding to general relativity with a cosmological constant. We start with the Plebanski formulation \cite{Plebanski:1977zz} of the theory, and then integrate out the two-form field, as well as the Lagrange multiplier field. A similar in spirit derivation is given in \cite{Capovilla:1991kx}. However, the final result of that calculation is erroneous, see \cite{CDJ-erratum}. Here we present the correct defining function. 

In the Plebanski formulation GR with a cosmological constant $\Lambda$ is described by the following action
\be
S[B,A,\Psi]=\frac{1}{8\pi \im G}\int \left[B^i\wedge F^i - \frac{1}{2} \left( \Psi^{ij}+ \frac{\Lambda}{3}\delta^{ij}\right) B^i\wedge B^j\right].
\ee
Here $G,\Lambda$ are the Newton's and cosmological constant respectively, $B^i$ is a ${\mathfrak su}(2)$-valued two-form field, $A^i$ is a ${\rm SU}(2)$ connection, $\im=\sqrt{-1}$, and $\Psi^{ij}$ is the symmetric traceless field of Lagrange multipliers. More details on this formulation can be found in e.g. \cite{Krasnov:2009pu}. Integrating out the two-form field one gets the following action
\be
S[A,\Psi] = \frac{1}{16\pi\im G} \int \left( \Psi^{ij}+ \frac{\Lambda}{3}\delta^{ij}\right)^{-1} F^i\wedge F^j,
\ee
where it is assumed that the matrix $\left( \Psi^{ij}+ (\Lambda/3)\delta^{ij}\right)$ is invertible. It is now convenient to rescale the Lagrange multipliers field and write the action as
\be
S[A,\tilde{\Psi}] = \frac{1}{\im} \int \left(\tilde{\Psi}^{ij} + \alpha \delta^{ij} \right)^{-1} F^i\wedge F^j,
\ee
where
\be
\alpha:= \frac{16\pi G\Lambda}{3}
\ee
is a dimensionless quantity. Note that $\alpha\sim M_0^2/M_p^2$ and so is of the order $\alpha\sim 10^{-120}$. 

In the final step we integrate out the Lagrange multiplier field $\tilde{\Psi}^{ij}$. Let us drop the tilde on the symbol for brevity. We can the rewrite the above action as
\be
S[A,\Psi]=\frac{1}{\im} \int ({\rm vol}) {\rm Tr}\left((\Psi +\alpha {\rm Id})^{-1} X\right),
\ee
where we have introduced $F^i\wedge F^j = ({\rm vol}) X^{ij}$, and $({\rm vol})$ is an arbitrary auxiliary 4-form on our manifold. To integrate out the matrix $\Psi$ we have to solve the field equations for it, and then substitute the result back into the action. Assuming that the solution for $\Psi$ can be written as a function of the matrix $X$ that admits a representation as a series in powers of $X$, we see that $\Psi$ will be diagonal if $X$ is. Thus, we can simplify the problem of finding $\Psi$ by using an ${\rm SO}(3)$ rotation to go to a basis in which $X$ is diagonal. This is always possible at least locally. We then look for a solution in which $\Psi$ is also diagonal. Denoting by $\lambda_1,\lambda_2,\lambda_3$ the eigenvalues of $X^{ij}$, and by $a,b,-(a+b)$ the components of the diagonal matrix $\Psi$, we get the following action functional to consider
\be\label{app-1}
F[a,b,\lambda]= \frac{\lambda_1}{\alpha+a}+\frac{\lambda_2}{\alpha+b} + \frac{\lambda_3}{\alpha-(a+b)}.
\ee
We now have to vary this with respect to $a,b$ and substitute the solution back to obtain the defining function as a function of $\lambda_i$. Assuming that neither of the denominators in (\ref{app-1}) is zero we get the following two equations
\be
(\alpha+a)^2 \lambda_3 = (\alpha-(a+b))^2\lambda_1, \qquad (\alpha+b)^2 \lambda_3 = (\alpha-(a+b))^2\lambda_2.
\ee
Taking the (positive branch of the) square root and adding the results we get $a+b$, which is most conveniently written as 
\be
\alpha-(a+b)=3\alpha \frac{\sqrt{\lambda_3}}{\sqrt{\lambda_1}+\sqrt{\lambda_2}+\sqrt{\lambda_3}}.
\ee
The other two combinations that appear in (\ref{app-1}) are given by similar expressions:
\be
\alpha+a = 3\alpha \frac{\sqrt{\lambda_1}}{\sqrt{\lambda_1}+\sqrt{\lambda_2}+\sqrt{\lambda_3}}, \qquad \alpha+b = 3\alpha \frac{\sqrt{\lambda_2}}{\sqrt{\lambda_1}+\sqrt{\lambda_2}+\sqrt{\lambda_3}}.
\ee
It is now clear that the defining function is
\be\label{def-GR}
f_{GR}(\lambda) = \frac{1}{3\alpha} \left( \sqrt{\lambda_1}+\sqrt{\lambda_2}+\sqrt{\lambda_3}\right)^2 = \frac{1}{3\alpha}\left({\rm Tr}\sqrt{X}\right)^2.
\ee

Thus, we learn that the action for GR with the cosmological constant can be rewritten in the form (\ref{action}) as follows
\be\label{GR}
S_{GR}[A] = \frac{1}{16\pi\im G\Lambda} \int \left( {\rm Tr}\sqrt{F^i\wedge F^j}\right)^2.
\ee

For the defining function (\ref{def-GR}) the value $f_0=f(\delta)$ is given by
\be
f_0 = \frac{3}{\alpha},
\ee
which is thus of the order $f_0\sim 10^{120}$. We can also compute the constant $g_0$. Thus, from (\ref{fpp}) we get
\be
\frac{\partial^2 f}{\partial \lambda_1 \partial \lambda_1} \Big|_{\lambda_i=1} = \frac{4g_0}{3}.
\ee
On the other hand, evaluating the second derivative of (\ref{def-GR}) with respect to $\lambda_1$ we get
\be
\frac{\partial^2 f_{GR}}{\partial \lambda_1 \partial \lambda_1} \Big|_{\lambda_i=1}= -\frac{1}{3\alpha}.
\ee
Thus, 
\be
g_0 = -\frac{1}{4\alpha},
\ee
where the minus sign reflects the concave character of (\ref{def-GR}). Thus we have $|g_0|\sim 10^{120}$ and $f_0/g_0=-12$ for this defining function.

\section*{Appendix B: Defining function for the (minimally) modified GR}

In this Appendix we analyse the defining function for what can be called minimally modified general relativity. The Plebanski-like action is given by:
\be
S[B,A,\Psi]=\frac{1}{8\pi \im G}\int \left[B^i\wedge F^i - \frac{1}{2} \left( \Psi^{ij}+ \frac{\Lambda}{3}\delta^{ij}+ \frac{\tilde{g}}{2}{\rm Tr}(\Psi^2) \delta^{ij}\right) B^i\wedge B^j\right].
\ee
Here $\tilde{g}$ is a constant of dimensions $\tilde{g}\sim M^{-2}$. Thus, this theory contains, in addition to $G, \Lambda$ present in GR, an additional dimensionful coupling $\tilde{g}$. As before, we now integrate out the two-form field, and then rescale all the quantities by a multiple of $16\pi G$. Thus, we introduce:
\be
g:= \frac{\tilde{g}}{16\pi G},
\ee
which is dimensionless, and then write the action omitting the tilde from over the symbol of $\Psi$. The resulting action is:
\be
S[A,\Psi]=\frac{1}{\im} \int ({\rm vol}) {\rm Tr}\left(\left(\Psi +\alpha {\rm Id}+ \frac{g}{2} {\rm Tr}(\Psi^2)  {\rm Id} \right)^{-1} X\right).
\ee
Since it is natural to expect that the scale of deformation set by $\tilde{g}$ is of the order of $\tilde{g}\sim M_p^{-2}$, the natural values for $g$ are order 1. The action then contains a small parameter $\alpha$, and the action with $\Psi$ integrated out can be found as an expansion in powers of this parameter. 

As in the previous section we will integrate out $\Psi$ by first diagonalising $X^{ij}$ and then looking for a solution for $\Psi^{ij}$ as a function of $X^{ij}$, which guarantees that it is also diagonal. Thus, we have to consider the following functional of the eigenvalues only:
\be\label{app-2}
F[a,b,\lambda]= \frac{\lambda_1}{\alpha+a+ g(a^2+b^2+ab)}+\frac{\lambda_2}{\alpha+b+ g(a^2+b^2+ab)} + \frac{\lambda_3}{\alpha-(a+b)+ g(a^2+b^2+ab)},
\ee
where as before $\lambda_1,\lambda_2,\lambda_3$ are eigenvalues of $X^{ij}$ and $a, b, -a-b$ are those of $\Psi^{ij}$. We now differentiate with respect to $a,b$ and get the following two equations:
\be
\frac{\lambda_1(1+g(2a+b))}{(\alpha+a+ g(a^2+b^2+ab))^2}= \frac{\lambda_3(1-g(2a+b))}{(\alpha-a-b+ g(a^2+b^2+ab))^2}, \\ \nonumber
\frac{\lambda_2(1+g(2b+a))}{(\alpha+b+ g(a^2+b^2+ab))^2}= \frac{\lambda_3(1-g(2b+a))}{(\alpha-a-b+ g(a^2+b^2+ab))^2}.
\ee
We now look for the solutions in the form of a series:
\be
a=a^{(1)}+a^{(2)}+\ldots, \qquad b=b^{(1)}+b^{(2)}+\ldots,
\ee
where $a^{(1)}, b^{(1)}$ is $O(\alpha)$, $a^{(2)}, b^{(2)}$ is $O(\alpha^2)$, and the dots denote higher orders in the small parameter $\alpha$. We have already found above that
\be
a^{(1)}= \alpha \frac{2\sqrt{\lambda_1} - \sqrt{\lambda_2}-\sqrt{\lambda_3}}{\sqrt{\lambda_1}+\sqrt{\lambda_2}+\sqrt{\lambda_3}}, \qquad b^{(1)}= \alpha \frac{2\sqrt{\lambda_2} - \sqrt{\lambda_1}-\sqrt{\lambda_3}}{\sqrt{\lambda_1}+\sqrt{\lambda_2}+\sqrt{\lambda_3}}.
\ee
Using this we get:
\be
(a^{(1)})^2+(b^{(1)})^2+a^{(1)} b^{(1)} = \frac{\alpha^2}{2(\sum_i\sqrt{\lambda_i})^2} \left(6\lambda_1+6\lambda_2+6\lambda_3-7\sqrt{\lambda_1\lambda_2} -7\sqrt{\lambda_1\lambda_3}-7\sqrt{\lambda_2\lambda_3}\right).
\ee
Let us introduce a compact notation
\be
\Delta:=6\lambda_1+6\lambda_2+6\lambda_3-7\sqrt{\lambda_1\lambda_2} -7\sqrt{\lambda_1\lambda_3}-7\sqrt{\lambda_2\lambda_3}.
\ee
The functional (\ref{app-2}) can the be rewritten as follows:
\be
F[a,b,\lambda]=\frac{\sum_i\sqrt{\lambda_i}}{3\alpha} \Big(\sqrt{\lambda_1}\left(1+\frac{(\sum_i\sqrt{\lambda_i})a^{(2)}}{3\alpha \sqrt{\lambda_1}} + \frac{\alpha g\Delta }{6 (\sum_i\sqrt{\lambda_i}) \sqrt{\lambda_1}}+O(\alpha^2)\right)^{-1}  \\ \nonumber +\sqrt{\lambda_2}\left(1+\frac{(\sum_i\sqrt{\lambda_i})b^{(2)}}{3\alpha \sqrt{\lambda_2}} + \frac{\alpha g\Delta }{6 (\sum_i\sqrt{\lambda_i}) \sqrt{\lambda_2}}+O(\alpha^2)\right)^{-1} \\ \nonumber +\sqrt{\lambda_3}\left(1-\frac{(\sum_i\sqrt{\lambda_i})(a^{(2)}+b^{(2)})}{3\alpha \sqrt{\lambda_3}} + \frac{\alpha g\Delta }{6 (\sum_i\sqrt{\lambda_i}) \sqrt{\lambda_3}}+O(\alpha^2)\right)^{-1} \Big).
\ee
Expanding the denominators in a power series in $\alpha$ and keeping only the $O(\alpha)$ terms we see that the terms involving $a^{(2)}, b^{(2)}$ cancel, and so we don't need to find these quantities to this order in $\alpha$. We get the following functional
\be\label{app-3}
F[\lambda]=\frac{(\sum_i\sqrt{\lambda_i})^2}{3\alpha} - \frac{g\Delta}{6} + O(\alpha).
\ee
We can rewrite it in a more convenient form by noting that the function
\be\label{app-top}
F_{\rm top}[\lambda]=\sum_i \lambda_i
\ee
gives rise to a total derivative, and so can always be added to our action. Thus, we can neglect multiples of $F_{\rm top}[\lambda]$. It is then easy to see that the function (\ref{app-3}) modulo (\ref{app-top}) is equal to
\be\label{app-4}
F[\lambda]\approx \frac{(\sum_i\sqrt{\lambda_i})^2}{3\alpha}\left(1+ \frac{7\alpha g}{4} + O((\alpha g)^2) \right),
\ee
where $\approx$ stands for equal modulo $F_{\rm top}[\lambda]$. The fact that it is the combination $\alpha g$ whose powers appear in brackets can be seen from (\ref{app-2}). Indeed, one can rescale the variables $a\to \alpha a, b\to \alpha b$ in (\ref{app-2}) so as to take $1/\alpha$ outside of the functional. Then the denominators will contain the combination $\alpha g$, and it is clear that the function with $a, b$ integrated out can be represented as an expansion in powers of $\alpha g$. The first term in this expansion is given in (\ref{app-4}).

\end{document}